\documentclass[12pt,hidelinks]{article}
\usepackage[utf8]{inputenc}
\usepackage{blindtext}

\usepackage[table]{xcolor}
\usepackage{easyReview}
\usepackage{derivative}
\usepackage{natbib}
\usepackage{bibunits}          
\defaultbibliographystyle{apalike} 
\defaultbibliography{CausalMissing}
\usepackage{amsmath,amsthm,amsfonts,amssymb,amscd,authblk,bbm,bm,booktabs,caption,floatrow,geometry,multirow,threeparttable,multicol,tikz,derivative,esdiff,subfig}
\usepackage{hyperref,url}
\hypersetup{
    colorlinks=false,
    linkcolor=blue,
    urlcolor=blue,
    citecolor=blue
}

\newtheorem{theorem}{Theorem}
\newtheorem{assumption}{Assumption}
\newtheorem{example}{Example}

\def\ipw{{\rm ipw}}
\def\reg{{\rm reg}}
\def\dr{{\rm dr}}
\def\expit{{\rm expit}}

\def\T{{\rm T}}

\def\spacingset#1{\renewcommand{\baselinestretch}%
{#1}\small\normalsize} \spacingset{1}

\title{\bf Correcting nonignorable nonresponse bias in turnout estimation using callback data}

\author[1]{Xinyu Li\textsuperscript{*}}
\author[1]{Naiwen Ying\textsuperscript{*}}
\author[2]{Kendrick Qijun Li}
\author[3]{Xu Shi}
\author[1]{Wang Miao\textsuperscript{\textdagger}}
  
\affil[1]{Department of Probability and Statistics, Peking University}
\affil[2]{Department of Biostatistics, St. Jude Children's Research Hospital}
\affil[3]{Department of Biostatistics, University of Michigan}

\date{}

\begin{document}
\maketitle

\renewcommand{\thefootnote}{*}
\footnotetext{Equal contribution.}
\renewcommand{\thefootnote}{\textdagger}
\footnotetext{Corresponding author: Wang Miao; Email: mwfy@pku.edu.cn}
\renewcommand{\thefootnote}{\arabic{footnote}}

\begin{abstract}
Overestimation of turnout has long been an issue in election surveys, with nonresponse bias or voter overrepresentation identified as major sources of bias.
However, adjusting for nonignorable nonresponse bias is substantially challenging.
Based on the ANES Non-Response Follow-Up study concerning the 2020 U.S. presidential election, we investigate the role of callback data, that is, records of contact attempts in the survey course, in adjusting for nonresponse bias in the estimation of turnout.
We propose a stableness of resistance assumption to account for nonignorable missingness in the outcome, which states that the impact of the missing outcome on the response propensity is stable in the first two call attempts.
Under this assumption and by integrating with covariate information from the census data, we establish identifiability and develop estimation methods for turnout.
Our methods produce estimates very close to the official turnout and successfully capture the trend of declining willingness to vote as response reluctance increases.
This work highlights the importance of adjusting for nonignorable nonresponse bias and demonstrates the potential of widely available callback data for political surveys.

\end{abstract}

\noindent%
\vfill

\spacingset{1.5}

\begin{bibunit}

\section{Introduction}\label{sec:intro}

Overestimation of voter turnout has been a persistent challenge for election surveys. Since the late 1940s, turnout estimates from the American National Election Study (ANES) have consistently exceeded official reports, often by more than 15 percentage points in recent presidential elections \citep{brehm2009phantom,burden2000voter,dahlgaard2019bias,enamorado2019validating,jackman2019does}. 
A key source of bias is the presence of nonignorable nonresponse bias: those who do not vote are less likely to respond to the election survey \citep{berent2011quality,sciarini2017lost}. 
In this situation, the missingness is typically related to the unobserved outcome itself, referred to as nonignorable or missingness not at random \citep[MNAR,][]{little2019statistical}, making valid adjustment substantially more challenging than under the missingness at random (MAR) assumption, where the missingness is independent of the missing outcome conditional on fully observed covariates.

Recent work has developed parametric and auxiliary variable-based approaches for MNAR adjustment \citep{heckman1979sample,liu2020identification,miao2016identifiability,miao2016varieties,sun2018semiparametric}. 
Notably, \citet{bailey2024polling,bailey2025countering} proposed a randomized response instrument (RRI) framework by using randomized treatments as response instruments to adjust for nonresponse bias, which is an important step in advancing political survey research toward MNAR-based analysis.
Alternatively, a promising direction is the use of callback data---records about the data collection process. In political surveys, callback data are particularly valuable, as they record the effort needed to obtain responses and are widely available \citep{biemer2013using,couper1998measuring,olson2013paradata}. 
With callback data, the continuum of resistance (COR) model approximates nonrespondents by the most reluctant respondents \citep{clarsen2021revisiting,lin1995using},
although, in certain situations, nonrespondents are still quite different from the hardest-to-reach respondents, limiting the model's ability to fully correct nonresponse bias in certain contexts.

This article introduces a stableness of resistance (SOR) model for leveraging callback data to adjust for nonignorable nonresponse and applies it to the ANES Non-Response Follow-Up (NRFU) survey concerning the 2020 U.S. presidential election to estimate the voter turnout.
It offers a promising complement to the RRI framework of \citet{bailey2025countering}.
The SOR model assumes that the impact of the missing outcome on the resistance or willingness to respond remains the same across the first two call attempts.
Similar ideas have been applied by \citet{alho1990adjusting,guan2018semiparametric,kim2014propensity,peress2010correcting,qin2014semiparametric} with a fully parametric propensity score model.
Building on a previous framework \citep{miao2025stableness}, 
this article adopts a semiparametric approach that only requires partially parametric specification of the joint distribution,
and extends the framework of \citet{miao2025stableness} to accommodate the joint missingness of both the outcome and covariates, 
using census data to recover the covariates distribution. 
We develop semiparametric estimators, including a doubly robust one that yields consistent estimates when either the second-call response propensity model or the outcome model is correctly specified, provided correct specification of the first-call response propensity model and an odds ratio model about the outcome-missingness relationship.

Our analysis produces turnout estimates closely aligned with official reports, provides strong evidence of nonignorable nonresponse in the NRFU survey and reveals systematic heterogeneity in survey participation and voting behavior. For example, women and senior people are more likely to respond, better-educated and senior people are more likely to vote, while Hispanics exhibit lower likelihoods of both responding and voting.
Besides, the association between design covariates and response may be heterogeneous across different demographic groups, for instance, visible cash incentives tend to increase response rates more among men than among women.
These findings highlight both the methodological potential of callback data and the practical implications for designing future political surveys.

\section{The ANES 2020 NRFU study}

The ANES 2020 NRFU study is a follow-up to the ANES 2020 Time Series Study, designed to collect data to analyze nonresponse bias, providing a valuable case study to explore self-reported turnout. 
The study was conducted by mail with 8,000 addresses from the ANES 2020 Time Series Study, constituting a weighted sample from the original ANES population.
It began on January 28, 2021, with an advance postcard randomly sent as part of a factorial design.
The study contains a two-stage callback design:
The first class invitation was mailed on February 1, followed by the second class invitation sent to nonrespondents in the first class with replacement questionnaire on March 2 and March 30.
This survey procedure provides us with high-quality callback data for nonresponse adjustment.
Besides, reminder postcards were sent on February 16 and April 5.

The NRFU study also embedded a factorial design for several methodological experiments to investigate the effects of mail-based design features on the response rate, concerning the study title, advance postcard, questionnaire length and content and visible cash incentives.
These methodological experiments are independently randomized.

After two callback stages, the NRFU survey closed on June 1, 2021, with 3,779 completed questionnaires.
The response rate is 48.3\% after excluding 168 undeliverable, deceased and removed cases from the total samples.
According to \citet{McDonald2024}, the official turnout in voting-age population (VAP) and voting-eligible population (VEP) in the 2020 presidential election is 62.0\% and 66.4\%, respectively.
Since the ANES sampling framework excludes noncitizens, the VEP rate can be viewed as an approximation to the true turnout of ANES population.
However, the weighted voter turnout based on the respondents in the NRFU study is over 85\%, which is much higher than the VEP or VAP turnout, indicating a severe nonresponse bias.

\section{Methodology}\label{sec:method}

\subsection{Model assumptions and identification}

In this section, we illustrate how to use callback data and external census data to identify the true turnout and adjust for nonresponse bias in the NRFU study, where both demographic covariates and outcomes are subject to missingness due to nonresponse.
Intuitively, callback data reflect the reluctance of units to respond and differences in turnout across callback stages suggest an association between voting and response propensity, indicating nonignorable nonresponse.
In the NRFU survey, voter turnout declines from 87.9\% among first-call respondents to 81.9\% among second-call respondents, and falls to 42.4\% among nonrespondents (based on VEP).
This declining trend provides empirical evidence that harder-to-reach individuals are less likely to vote, motivating our modeling strategy to adjust for such nonignorable nonresponse using callback data.

We now formalize this idea.
Let $Y$ denote the voting behavior, with $Y=1$ indicating that an individual voted and $Y=0$ otherwise.
We are interested in the outcome mean $E(Y)$, that is, the voter turnout in the ANES population.
Let $X$ denote a vector of covariates collected in the survey, possibly including components $X_1$ (e.g., age and gender) that are missing together with $Y$ and also components $X_2$ that are fully observed.
Note that $X_2$ could be an empty set.
In the NRFU study, $(X_1,Y)$ are missing for nonrespondents, and $X_2$ is a vector related to the factorial design experiments, for example, questionnaire version and whether the prepaid cash incentives are visible.
The experimental design is randomized, that is, $X_2$ is independent of $X_1$.
To simplify the notation, we no longer isolate demographic covariates or design covariates from $X$ unless otherwise specified.
Follow-ups are conducted to increase the response rate.
Let $R_k$ ($k = 1, \ldots, K$) denote the callback data, where $R_k = 1$ if an individual's answers are available at the $k$th contact and $R_k = 0$ otherwise.
By definition, respondents with $R_{k+1}=1$ include those who respond before the $k+1$th call and those who respond in the $k+1$th call.
Let $f$ denote a generic probability density or mass function, $f(X,Y,R_1,\ldots,R_K)$ denote the unknown full-data distribution of interest, and $\mathcal{O}=\{R_K X_1,X_2, R_K Y,R_1,\cdots,R_K\}$ denote the observed variables, including the questionnaire data and callback data.
Table~\ref{tbl:1} illustrates the data structure of a survey with callbacks.
The observed data consist of $n$ independent samples of $\mathcal{O}$ with known sampling weights $w_i$ attached to the $i$th unit, $i=1,\ldots,n$.
For notational convenience, we use capital letters for random variables and lowercase letters for their realized values.
In the NRFU study, adjusting for the sampling weights ensures that the NRFU sample is representative of the ANES population.

 \begin{table}[!ht] 
 \centering
 \begin{threeparttable}
\caption{Data structure of a survey with callbacks}\label{tbl:1}
\begin{tabular}{cccccccccccc}
\toprule    
& \multicolumn{2}{c}{Questionnaire data}   & &\multicolumn{5}{c}{Paradata}\\
\cmidrule{2-3}\cmidrule{5-9}
& turnout & demographic covariates & &\multicolumn{1}{c}{design covariates}&\multicolumn{4}{c}{callbacks}\\ 
\cmidrule{6-9}
ID&$Y$&$X_1$&& $X_2$ & $R_1$&$R_2$&$\ldots$&$R_K$\\
\midrule
1&$y_1$ & $x_{1,1}$ & &$x_{2,1}$ &1&1&$\cdots$&1\\
2&$y_2$ & $x_{1,2}$& &$x_{2,2}$ &0&1&$\cdots$&1\\
3&NA  & NA && $x_{2,3}$&  0&0&$\cdots$&0\\
$\colon$&NA & NA && $\colon$&  0&0&$\cdots$&0\\
$n$&$y_{n}$  &$x_{1,n}$&& $x_{2,n}$&  0&0&$\cdots$&1\\
\bottomrule
\end{tabular}
\begin{tablenotes}[hang]
    \item[]Note: NA stands for missing values.
\end{tablenotes}
\end{threeparttable}
\end{table}

Let $\pi_1(X,Y)=f(R_1=1\mid X,Y)$ and $\pi_k(X,Y)=f(R_k=1\mid R_{k-1}=0, X,Y)$ for $k=2,\cdots,K$ denote the response propensity scores for each call attempt.
Without loss of generality, the propensity score can be written as
\begin{eqnarray}\label{eq:1}
 \pi_k(X,Y)=\operatorname{expit}\{A_k(X)+\Gamma_k(X,Y)\},
\end{eqnarray}
where $A_k(x)=\operatorname{logit} \{\pi_k(X=x,Y=0)\}$ is referred to as the baseline propensity score and 
$$\Gamma_k(x,y)=\log \frac{\pi_k(X=x,Y=y) \{1-\pi_k(X=x,Y=0)\}}{\{1-\pi_k(X=x,Y=y)\}\pi_k(X=x,Y=0) },$$ 
is the log odds ratio function for the propensity score $\pi_k$ with $\Gamma_k(x,y=0)=0$.
The function $\Gamma_k(x,y)$ quantifies the log-scale change in the odds of response probability caused by a shift of $Y$ from the reference value $Y=0$ to a level $y$ while the covariates are controlled at a fixed value $X=x$.
By this definition, $\Gamma_k(x,y)$ can be viewed as a measure of the resistance/willingness to respond caused by the outcome, that is, the degree of nonignorable missingness.
If $\Gamma_k(x,y)=0$ for all possible $(x,y)$, the missingness mechanism reduces to MAR, in which case the outcome is not associated with the response probability after conditioning on covariates $X$.
The function $\Gamma_k(x,y)$ can be equivalently defined with respect to the conditional distribution of the outcome and covariates on the response status, which characterizes the difference of units across different response patterns
\begin{eqnarray*}
\Gamma_k(x,y)&=&\log \frac{f(Y=y, X=x\mid R_{k-1}=0, R_k=1) f(Y=0, X=x\mid R_{k-1}=0,R_k=0)}{f(Y=y, X=x\mid R_{k-1}=0,R_k=0) f(Y=0, X=x\mid R_{k-1}=0, R_k=1)}.
\end{eqnarray*}
The log odds ratio function is widely used in the missing data literature for characterizing the association between the missing outcome and response propensity; see \citet{chen2007semiparametric,kim2011semiparametric,miao2025stableness} for detailed discussions.

Without additional assumptions, the propensity scores and the full-data distribution are not identified under MNAR.
To achieve identification, we make the following SOR assumption.

\begin{assumption}[Stableness of resistance]\label{assump1}
 $\Gamma_1(X,Y)=\Gamma_2(X,Y)$ almost surely.
\end{assumption}

Assumption~\ref{assump1} states that, conditional on the covariates, the impact of the voting behavior on the willingness to respond remains the same across the first two contact attempts.
Nonetheless, it admits heterogeneous associations between covariates $X$ and response propensity across different calls.
The assumption imposes restrictions only on the propensity scores of the first two calls, leaving the propensity scores and outcome distributions for later call attempts unrestricted.
The assumption imposes no parametric models on the propensity scores or on the effects of covariates, which thus accommodates flexible modeling strategies for estimation.
In the special case where $\pi_k$ follows a linear logistic model, Assumption~\ref{assump1} has a more explicit form.

\begin{example}\label{exam1}
Assuming $\pi_k(X,Y)=\operatorname{expit}(\alpha_{k0}+\alpha_{k1}^\T X+\gamma_kY)$ for the propensity score model,  
it is a special case of Equation \eqref{eq:1} with $A_k(X)=\alpha_{k0}+\alpha_{k1}^\T X$ and $\Gamma_k(X,Y)=\gamma_kY$.
In this case, Assumption~\ref{assump1} is equivalent to that $\gamma_1=\gamma_2$, that is, equal odds ratio parameters.
However, coefficients of covariates $\alpha_{k1}$ can be arbitrary and vary across different calls.
\end{example}

Similar to various missing data problems, Assumption~\ref{assump1} is untestable based on observed data.
Thus, its validity should be justified based on domain-specific knowledge and needs to be investigated on a case-by-case basis.
To further illustrate the intuition behind the stableness assumption, in Section~\ref{ssec:s1.1} of the Supplementary Material we provide a generative model about how a unit makes choice to vote and decides to respond in the election survey, which is motivated from the discrete choice theory \citep{mcfadden2001economic}.
For example, a unit's interest in politics or the local political climate influences both the voting behavior and the willingness to respond to election surveys, which is unobserved and thus constitutes an important source of nonignorable nonresponse.
The stableness assumption is a good approximation if such unobserved characteristics do not dramatically change in a short period of time, that is, in two adjacent calls.
In this case, we may expect the relationship between voting and response to be stable in two adjacent calls.
However, if the interest in politics or the local political climate dramatically changes, or the survey spans a long time range, the relationship between the voting behavior and the response may shift during the survey course and thus violate Assumption~\ref{assump1}.
As an anonymous reviewer commented, the assumption may also be violated by underlying heterogeneity in population.
Consider two types of respondents: those who always respond to the first call and those who do so probabilistically based on their characteristics.
If the proportion of ``always responders'' is small, the stableness assumption could still be a good approximation, but otherwise, conditioning on first-call nonresponse may yield a systematically different subpopulation, potentially violating this assumption.
These issues of heterogeneous unmeasured factors or population strata could be partially addressed by collecting richer measures of political orientation and prosocial behavior.
In addition, sensitivity analysis is warranted to assess the robustness of inference to potential violation of Assumption~\ref{assump1}; see Section~\ref{ssec:5.2}.

If covariates are fully observed, \citet{miao2025stableness} have shown that the full-data distribution is identified under a stableness assumption for the missing outcome; otherwise, a much stronger assumption covering the joint vector of both the missing outcome and missing covariates is required for identification.
However, a key difference in our setting here is that we only require the stableness assumption for the missing outcome (i.e., the voting behavior) and have no restriction on that of the missing covariates, because assuming stable influence of the demographic covariates could be fairly stringent in the election survey.
Therefore, the identification results by \citet{miao2025stableness} are not directly applicable here.
Instead, we consider the following identification strategy.

\begin{theorem}\label{theorem1}
Under Assumption~\ref{assump1}, and suppose that  $0<\pi_k(X,Y)<1$ for $k=1,2$, 
then the full-data distribution $f(X, Y, R_1, \cdots,R_K)$ is identified from the observed-data distribution $f(\mathcal{O})$ and the covariates distribution $f(X)$.
\end{theorem}

The proof of Theorem~\ref{theorem1} is provided in Section~\ref{ssec:s3.1} of the Supplementary Material, and a parameter-counting illustration of the identification result is provided in Section~\ref{ssec:s1.2} of the Supplementary Material.
To the best of our knowledge, Theorem~\ref{theorem1} is so far the most general result for identification with callback data.
Assumption~\ref{assump1} is a minimal assumption for identification and is considerably more flexible than previous proposals \citep[e.g.,][]{alho1990adjusting,guan2018semiparametric,kim2014propensity} that rely on strong parametric models to achieve identification.
Note that when multiple calls $(K\geq3)$ are available, identification of the joint distribution $f(X,Y,R_1, \ldots,R_K)$ equally holds even if the stableness assumption only holds for the first two calls.
Theorem~\ref{theorem1} complements the identification strategy of \citet{miao2025stableness} by incorporating covariates information from census data while only requiring stableness for the missing outcome.
In the NRFU study, we obtain the distribution $f(X_1)$ of missing demographic covariates from the census data and the distribution $f(X_2)$ of randomized design covariates is available directly from the survey.
Consequently, the joint distribution of covariates $f(X)=f(X_1)f(X_2)$ can be obtained as their product.
The propensity scores $\pi_k(X,Y)$ for the first two calls are assumed to be bounded away from zero and one, which is a regularity condition often entailed in missing data analysis.
The identification result does not require additional restriction on the functional form of the propensity score or on the outcome distribution.
Based on this identification result, we next propose feasible modeling and estimation methods for inference about the voter turnout.

\subsection{Estimation methods}\label{ssec:esti}
Let $\theta$ denote a generic parameter of interest, which is defined as the unique solution to a given full-data estimating equation $E\{m(X, Y ; \theta)\} = 0$,
where $E(\cdot)$ denotes the expectation evaluated in the ANES population.
In particular, $m(x, y ; \theta)=y-\theta$ when $\theta$ represents the voter turnout, and $m(x, y ; \theta)=x\{y-\operatorname{expit}(x^\T\theta)\}$ when $\theta$ characterizes the association between the covariates and the voting behavior under a logistic outcome model.
Recall that $w_i$ is the sampling weight associated with unit $i$, if there were no missing data in the NRFU sample, then a consistent estimator of $\theta$ could be obtained by solving $\sum_{i=1}^{n}w_i m(x_i,y_i;\theta)=0$.
In the presence of missing data, Theorem~\ref{theorem1} shows that under the SOR assumption, it suffices to identify $\theta$ with a two-stage callback design and the aid of demographic covariates distribution available from the census data.
For simplicity, the following focuses on estimation under two callback attempts, with extensions to multiple calls provided in Section~\ref{sec:multi-callback} of the Supplementary Material.

\subsubsection{Inverse probability weighting}
For the ease of interpretation, we specify parametric working models for propensity scores $\pi_k(x,y)$ while leaving the outcome distribution unrestricted.
From Equation \eqref{eq:1}, this is equivalent to specifying parametric working models $A_k(x;\alpha_k)$ for the baseline propensity scores and $\Gamma(x,y;\gamma)$ for the log odds ratio function, respectively.
For notational simplicity, we denote $\pi_{k,i}(\alpha_k,\gamma)=\pi_k(x_i,y_i;\alpha_k,\gamma)$.

We first obtain estimators $(\hat \alpha_{1,\ipw},\hat \alpha_{2,\ipw},  \hat \gamma_{\ipw})$ of the nuisance parameters $(\alpha_1,\alpha_2,\gamma)$ by solving the following estimating equations:
\begin{eqnarray}
E_{{f}}\{V_1(X)\}&=& \sum\limits_{i=1}^n w_i\left\{\frac{r_{1,i}}{\pi_{1,i}(\hat \alpha_{1,\ipw},\hat \gamma_{\ipw})}  \cdot V_1(x_i)\right\},\label{eq:3}\\
E_{{f}}\{V_2(X)\}&=& \sum\limits_{i=1}^n w_i \left[\left\{\frac{r_{2,i}-r_{1,i}}{\pi_{2,i}(\hat \alpha_{2,\ipw},\hat \gamma_{\ipw})}+r_{1,i}\right\}\cdot V_2(x_i) \right],\label{eq:4}\\
0&=& \sum\limits_{i=1}^n w_i \left[\left\{\frac{r_{2,i} - r_{1,i}}{ \pi_{2,i}(\hat\alpha_{2,\ipw},\hat \gamma_{\ipw})}  -  \frac{1-\pi_{1,i}(\hat\alpha_{1,\ipw},\hat \gamma_{\ipw})}{\pi_{1,i}(\hat \alpha_{1,\ipw},\hat \gamma_{\ipw})}r_{1,i} \right\}\cdot U(x_i,y_i)\right].\label{eq:5}
\end{eqnarray}
Equations \eqref{eq:3} and \eqref{eq:4} are constructed by calibrating the sample mean of covariates functions in the first- and second-call respondents to that of the target population, respectively.
Here $V_1(x) = \partial A_1(x; \alpha_1)/\partial \alpha_1$, $V_2(x) = \partial A_2(x; \alpha_2)/\partial \alpha_2$,
$U(x, y) =\partial \Gamma(x, y; \gamma)/\partial \gamma$.
For instance, under a linear logistic model with $A_1(x; \alpha_1) =x^\T\alpha_1$, $A_2(x; \alpha_2) =x^\T\alpha_2$, $\Gamma(x, y; \gamma) = y\gamma$, one may use $V_1(x) = V_2(x) = x$, $U(x, y) =y$.
Note that $V_1, V_2, U$ can be chosen as other user-specified vector functions; see \citet[][page 30]{tsiatis2006semiparametric} for a general recommendation.
The expectation $E_{{f}}\{V_k(X)\}=\int V_k(x){f}(x)dx$ is calculated according to the distribution of covariates in the target population, with $ f(x)$ being the distribution of $X$ obtained from the observed data and census data.

Given nuisance estimators $(\hat \alpha_{1,\ipw},\hat \alpha_{2,\ipw},  \hat \gamma_{\ipw})$,
an inverse probability weighted (IPW) estimator of $\theta$ is the solution $\hat\theta_{\ipw}$ to the following equation: 
\begin{equation}\label{eq:6}
    0  = \sum\limits_{i=1}^n w_i\left[  \frac{r_{2,i}\cdot m(x_i,y_i;\hat\theta_{\ipw})}{\pi_{1,i}(\hat\alpha_{1,\ipw},\hat \gamma_{\ipw}) +  \pi_{2,i} (\hat\alpha_{2,\ipw},\hat \gamma_{\ipw}) \{1-  \pi_{1,i} (\hat\alpha_{1,\ipw},\hat \gamma_{\ipw}) \}} \right].
\end{equation}
Equations \eqref{eq:3}--\eqref{eq:6} can be jointly solved using the generalized method of moments \citep[][]{hansen1982large}. 

\subsubsection{Imputation/regression-based estimation}

An alternative estimation strategy is to impute the missing values in the construction of estimating equations, which entails parametric working models $\{f_2(y\mid x;\beta),\pi_1(x,y;\alpha_1,\gamma)\}$ for  
the second-call conditional outcome distribution $f_2(y\mid x)=f(y\mid x, r_2=1,r_1=0)$ and the first-call propensity score $\pi_1(x,y)$.
We estimate the nuisance parameters $(\alpha_1,\gamma, \beta)$ by solving
\begin{eqnarray}
0&=&\sum\limits_{i=1}^n w_i \left\{  (r_{2,i}-r_{1,i}) \cdot \left. \frac{ \partial \log f_2(y_i\mid x_i;\beta)}{\partial \beta}\right|_{\beta=\hat \beta_{\reg}} \right\},\label{eq:7} \\
E_{ f}\{h_U(X;\hat{\beta}_{\reg},\hat{\gamma}_{\reg})\}&=&\sum\limits_{i=1}^n w_i\left[  
\begin{aligned}
\left\{\frac{r_{1,i}}{\pi_{1,i}(\hat \alpha_{1,\reg},\hat \gamma_{\reg})} -r_{2,i}\right\} U(x_i,y_i)\\
+  r_{2,i} h_U(x_i;\hat\beta_{\reg},\hat \gamma_{\reg})  
\end{aligned}
\right],\label{eq:8}
\end{eqnarray} 
where $U(x,y)=\{\partial A_1(x; \alpha_1)/\partial \alpha_1,\partial \Gamma(x, y; \gamma)/\partial \gamma\}$.
Note that $U$ can be chosen as other user-specified vector functions; see \citet[][page 30]{tsiatis2006semiparametric}.
The imputation equation \eqref{eq:8} is motivated by the following equation:
\begin{eqnarray*}
    f(Y\mid X,R_2=0)&=&\frac{\exp(-\Gamma)f(Y\mid X,R_2=1,R_1=0)}{E\{\exp(-\Gamma)\mid R_2=1,R_1=0\}},
\end{eqnarray*}
which recovers the conditional outcome distribution in the nonrespondents with the observed-data distribution and the odds ratio function.
This equation, known as exponential tilting, has been widely used in the analysis of nonignorable missing data; see \citet{bailey2025countering,kim2011semiparametric,miao2025stableness,sun2018semiparametric} for examples.
Based on this equation, we can impute the missing values of any function $U(X,Y)$ that involves the missing outcome, with
\begin{equation*}
h_U(x;\beta,\gamma)=E\{U(X,Y) \mid X=x, R_2=0;\beta,\gamma\}
= \int U(x,y) \frac{\exp\{-\Gamma(x,y;\gamma)\}f_2(y\mid x;\beta)}{\int \exp\{-\Gamma(x,y;\gamma)\}f_2(y\mid x;\beta) dy} dy.
\end{equation*}
Then we solve the following estimating equation to obtain the estimator $\hat\theta_{\reg}$:
\begin{equation}\label{eq:9}
    0  = \sum\limits_{i=1}^n w_i\left[  r_{2,i} \Big\{m(x_i,y_i;\hat\theta_{\reg})-h_m(x_i;\hat\theta_{\reg},\hat\beta_{\reg},\hat\gamma_{\reg}) \Big\}\right]+E_{ f}\Big\{h_m(X;\hat\theta_{\reg},\hat\beta_{\reg},\hat\gamma_{\reg})\Big\},
\end{equation}
where $h_m(x;\theta,\beta,\gamma)=E\{m(X,Y;\theta) \mid X=x, R_2=0;\beta,\gamma\}$ imputes the missing values of $m(X,Y;\theta)$.
We call $\hat \theta_{\reg}$ an imputation/regression-based (REG) estimator.
The covariates distribution $ f(x)$ obtained from the observed data and census data is involved in the expectation $E_{ f}$ to calibrate the estimation of $(\alpha_1, \gamma)$.

Under the conditions of Theorem~\ref{theorem1}, a stronger positivity assumption that $c<\pi_k(X,Y)<1-c$, $k=1,2$ for some constant $0<c<1$, and regularity conditions described by \citet[][Theorems 2.6 and 3.4]{newey1994large},
the IPW estimators $(\hat\theta_{\ipw},\hat \alpha_{1, \ipw},\hat \alpha_{2,\ipw},\hat \gamma_{\ipw})$ are consistent and asymptotically normal if the propensity score models $\pi_1 (x, y;\alpha_1,\gamma )$ and $\pi_2 (x, y;\alpha_2,\gamma )$ are correctly specified,
and the REG estimators $(\hat\theta_{\reg},\hat \alpha_{1, \reg}, \hat \gamma_{\reg}, \hat \beta_{\reg})$ are consistent and asymptotically normal if $\pi_1 (x, y;\alpha_1,\gamma )$ and $f_2 (y\mid x,\beta)$ are correctly specified.

\subsubsection{Doubly robust estimation}\label{ssec:dr}
If the required propensity score or outcome distribution model is misspecified, the corresponding IPW or REG estimator is no longer consistent.
It is thus desirable to develop a doubly robust estimator that remains consistent even if one of the working models is misspecified, without knowing which is misspecified.
Such estimators have become increasingly popular in recent years for missing data analysis, causal inference and other problems with data coarsening \citep{bang2005doubly,tsiatis2006semiparametric}.

We specify working models $\{\pi_1(x,y;\alpha_1,\gamma),\pi_2(x,y;\alpha_2,\gamma), f_2(y\mid x;\beta)\}$ and
solve the following equations to obtain the nuisance estimators $(\hat \alpha_{1,\dr},\hat \alpha_{2,\dr}, \hat \beta_{\dr}, \hat \gamma_{\dr})$:
\begin{eqnarray}
E_{ f}\{V_1(X)\}&=& \sum\limits_{i=1}^n w_i\left\{\frac{r_{1,i}}{\pi_{1,i}(\hat \alpha_{1,\dr},\hat \gamma_{\dr})}  \cdot V_1(x_i)\right\},\label{eq:10}\\
E_{ f}\{V_2(X)\}&=& \sum\limits_{i=1}^n w_i \left[\left\{\frac{r_{2,i}-r_{1,i}}{\pi_{2,i}(\hat \alpha_{2,\dr},\hat \gamma_{\dr})}+r_{1,i}\right\}\cdot V_2(x_i) \right],\label{eq:11}\\
0 &=& \sum\limits_{i=1}^n w_i \left\{  (r_{2,i}-r_{1,i}) \cdot \left.\frac{ \partial \log f_2(y_i\mid x_i;\beta)}{\partial \beta}\right|_{\beta=\hat \beta_{\dr}} \right\},\label{eq:12}\\
0&=&\sum\limits_{i=1}^n w_i \left[
\begin{aligned}
\left\{r_{1,i} - \frac{\pi_{1,i}(\hat\alpha_{1,\dr},\hat\gamma_{\dr})}{1-\pi_{1,i}(\hat\alpha_{1,\dr},\hat\gamma_{\dr})}\frac{r_{2,i} - r_{1,i}}{ \pi_{2,i}(\hat\alpha_{2,\dr},\hat\gamma_{\dr})}  \right\}\\
\cdot \left\{ U(x_i,y_i) -  g_U(x_i;\hat \beta_{\dr}) \right\}
\end{aligned}
\right],\label{eq:13}
\end{eqnarray}
where $V_1(x) = \partial A_1(x; \alpha_1)/\partial \alpha_1$, $V_2(x) = \partial A_2(x; \alpha_2)/\partial \alpha_2$,
$U(x, y) =\partial \Gamma(x, y; \gamma)/\partial \gamma$, and $g_U(x;\beta)=E\left\{U(X,Y)\mid X=x, R_2=1,R_1=0;\beta\right\}=\int U(x,y)f_2(y\mid x;\beta)dy$.
Note that $U$ can be chosen as other user-specified vector functions; see \citet[][page 30]{tsiatis2006semiparametric}.
Then the doubly robust estimator $\hat \theta_{\dr}$ is obtained by solving
\begin{eqnarray}
    0 & =& \sum\limits_{i=1}^n w_i\left[ \left\{ r_{1,i} + \frac{r_{2,i}-r_{1,i}}{\pi_{2,i}(\hat \alpha_{2,\dr},\hat\gamma_{\dr})}\right\}\{m(x_i,y_i;\hat\theta_{\dr})-h_m(x_i;\hat\theta_{\dr},\hat\beta_{\dr},\hat\gamma_{\dr})\} \right]\nonumber\\
    &&+E_{ f}\Big\{h_m(X;\hat\theta_{\dr},\hat\beta_{\dr},\hat\gamma_{\dr})\Big\}.\label{eq:14}
\end{eqnarray}

Under the conditions of Theorem~\ref{theorem1}, a stronger positivity assumption that $c<\pi_k(X,Y)<1-c$, $k=1,2$ for some constant $0<c<1$, and regularity conditions described by \citet[][Theorems 2.6 and 3.4]{newey1994large},
estimators $(\hat \alpha_{1,\dr}, \hat \gamma_{\dr}, \hat \theta_{\dr})$ are consistent and asymptotically normal provided either one of the following conditions holds:
\begin{itemize}
  \item $A_1(x;\alpha_1)$, $\Gamma(x,y;\gamma)$, and $A_2(x;\alpha_2)$ are correctly specified; or
    \item $A_1(x;\alpha_1)$, $\Gamma(x,y;\gamma)$, and $f_2(y\mid x;\beta)$ are correctly specified.
\end{itemize}

Compared to $\hat \theta_{\ipw}$ and $\hat \theta_{\reg}$, the doubly robust estimator $\hat \theta_{\dr}$ offers one more chance to correct the bias due to model misspecification of $A_2(x;\alpha_2)$ or $f_2(y\mid x;\beta)$.
Note that $\Gamma(x,y;\gamma)$ and $A_1(x;\alpha_1)$, that is, the first-call propensity score model $\pi_1(x,y;\alpha_1,\gamma)$, needs to be correctly specified for the doubly robust estimator.

In practice, a political survey may involve multiple callbacks.
In Section~\ref{sec:multi-callback} of the Supplementary Material, we illustrate how to construct estimators that make use of data from all calls.
In addition, identification can be also achieved if Assumption~\ref{assump1} holds for any other two given adjacent calls, by viewing the $k$th and $k+1$th calls as the first two calls in a subsurvey on nonrespondents from the $k-1$th call (i.e., $R_{k-1}=0$).
In this spirit, it is viable to conduct robustness checks by taking different combinations of the callbacks for the stableness assumption.
Alternatively, assuming stableness across all calls will lead to more efficient estimation because the model class or number of parameters becomes much smaller.
However, this assumption could be overly restrictive and lead to less robust inference.

\section{Simulation study}\label{sec:simu}

We evaluate the performance of the proposed estimators via simulation studies.
Here, we consider a binary outcome and simulations for continuous outcome settings are presented in Section~\ref{ssec:simu-con} of the Supplementary Material.
Let $X = (1,X_a,X_b)^\T$ and $\widetilde{X} = (1,X_a^2,X_b^2)^\T$
with $X_a,X_b$ independent following a uniform distribution $\operatorname{Unif}(-1,1)$.
Table~\ref{tbl:2} presents settings for data generation and estimation in the simulation.
The samples of $(X,Y)$ are dropped for $R_2=0$ and the marginal distribution of $X$ is given for estimation.

\begin{table}[!ht]
\caption{Models for data generation and estimation in the simulation}
    \label{tbl:2}
    \centering
    \begin{tabular}{cccccccc}
    \toprule
  &  &\multicolumn{4}{c}{Data generating model for binary outcome}\\
  \cmidrule{3-6}
  &  &\multicolumn{4}{c}{$\pi_1=\expit(\alpha_1^\T X+\gamma Y), \pi_2=\expit(\alpha_2^\T W_1+\gamma Y),f_2(Y=1\mid X)=\expit(\beta^\T W_2)$}\\
   \cmidrule{3-6}
  &  &\multicolumn{4}{c}{Four scenarios with different choices of $(W_1, W_2)$}\\
  \midrule
    & Settings & TT & FT & TF & FF \\ 
    \midrule
   & $(W_1,W_2)$ & $(X,X)$ & $(\widetilde{X},X)$ & $(X,\widetilde{X})$ & $(\widetilde{X},\widetilde{X})$\\
& $\alpha_1^\T$ & $(-1, 0.5, 0.2)$ & $(-0.2, -0.5, 0.7)$ & $(-1, 0.5, 0.2)$ & $(-0.3, 0.5, 0.2)$ \\
& $\alpha_2^\T$ & $(-0.5, 0.5, 0.2)$ & $(-0.6, 1.7, 1.0)$ & $(-0.5, 0.5, 0.2)$ & $(-0.5, -1.5, 0.2)$ \\
& $\beta^\T$ & $(-0.5, 0.5, 0.5)$ & $(1.2, 0.5, 0.5)$ & $(-0.5, 5, -2.0)$ & $(-1, 5, 0.5)$ \\
& $\gamma$ & $1$ & $-0.9$ & $1.3$ & $1.5$ \\
&  & \multicolumn{4}{c}{Working model for estimation}\\
&  & \multicolumn{4}{c}{$\pi_1=\expit(\alpha_1^\T X+\gamma Y), \pi_2=\expit(\alpha_2^\T X+\gamma Y),f_2(Y=1\mid X)=\expit(\beta^\T X)$}\\
\bottomrule
    \end{tabular}
\end{table}

All working models are correctly specified in Scenario (TT). In Scenarios (TF) and (FF), the working model for the second-call outcome model is misspecified, and in Scenarios (FT) and (FF), the working model for the second-call baseline propensity score model is misspecified.
We implement the proposed IPW, REG, and DR methods to estimate the outcome mean, that is, the solution to $E\{m(X,Y;\theta)\}=E(Y-\theta)=0$.

We simulate 1,000 replicates for each scenario with a sample size of 5,000 in each replicate.
We set the functions $V_1(x)=V_2(x)=x$, and $U(x,y)=y$ in \eqref{eq:3}--\eqref{eq:5} for the IPW estimator; $U(x,y)=(x^\T,y)^\T$ for the REG estimator; $V_1(x)=V_2(x)=x$, and $U(x,y)=y$ for the DR estimator.
For comparison, we also implement standard estimators $(\hat \theta_{\ipw}^{\mathrm{mar}},\hat \theta_{\reg}^{\mathrm{mar}},\hat \theta_{\dr}^{\mathrm{mar}})$ that are IPW, REG and DR analogs based on MAR, with the number of callbacks included as an additional covariate, and all covariates are fully available.
The simulation results are summarized with boxplots of estimation bias in Figure~\ref{fig:1} for the outcome mean $\theta$ and Figure~\ref{fig:2} for the odds ratio parameter $\gamma$.

As expected, the three proposed estimators have little bias in Scenario (TT) where all working models are correctly specified.
However, the IPW and REG estimators have substantial bias when the second-call baseline propensity score model $A_2(x;\alpha_2)$ and the second-call outcome model $f_2(y\mid x;\beta)$ are misspecified, respectively.
In contrast, the DR estimator continues to exhibit little bias in these scenarios.
These results demonstrate the double robustness of $(\hat \theta_{\dr}, \hat \gamma_{\dr})$ against misspecification of either $A_2(x;\alpha_2)$ or $f_2(y\mid x;\beta)$.
In Scenario (FF), where both $A_2(x;\alpha_2)$ and $f_2(y\mid x;\beta)$ are misspecified, all three proposed estimators lead to biased estimates.
Besides, the three standard MAR estimators have large bias in all four scenarios, even if the number of callbacks is included as a covariate and all covariates are fully available.

We compute the variance of the proposed estimators and construct 95\% confidence intervals based on the normal approximation of their distributions.
We then assess the coverage rate of the confidence intervals and summarize the results in Table~\ref{tbl:3}.
In Scenarios (TT), (FT) and (TF), the 95\% confidence interval based on the DR estimator has a coverage rate that is very close to the nominal level of 0.95.
However, when the corresponding working model being misspecified, the 95\% confidence intervals based on the IPW and REG estimators have undersized coverage rates.

\begin{figure}[!ht]
\graphicspath{{figures/Figure_1/}}
\centering
\includegraphics[scale=0.45]{TT.pdf}
\includegraphics[scale=0.45]{FT.pdf}
\includegraphics[scale=0.45]{TF.pdf}
\includegraphics[scale=0.45]{FF.pdf}
\caption{Bias of estimators of $\theta$ in the binary outcome simulation.}
\floatfoot{\footnotesize
Note: Model $A_2$ is correctly specified in Scenarios (TT, TF), $f_2$ is correctly specified in Scenarios (TT, FT) and they are both misspecified in Scenario (FF).}
\label{fig:1}
\end{figure}

\begin{figure}[!ht]
\graphicspath{{figures/Figure_2/}}
\centering
\includegraphics[scale=0.45]{TT.pdf}
\includegraphics[scale=0.45]{FT.pdf}
\includegraphics[scale=0.45]{TF.pdf}
\includegraphics[scale=0.45]{FF.pdf}
\caption{Bias of estimators of $\gamma$ in the binary outcome simulation.}
\floatfoot{\footnotesize
Note: Model $A_2$ is correctly specified in Scenarios (TT, TF), $f_2$ is correctly specified in Scenarios (TT, FT) and they are both misspecified in Scenario (FF).}
\label{fig:2}
\end{figure}

\begin{table}[!ht]
\caption{Coverage rate of $95\%$ confidence interval in the binary outcome simulation}
\label{tbl:3}
\centering
\begin{tabular}{ccccccccccc}
\toprule
& \multicolumn{6}{c}{$\theta$} && \multicolumn{3}{c}{$\gamma$}\\ 
\cmidrule{2-7}\cmidrule{9-11}
Scenarios & IPW & REG & DR & IPW$_\mathrm{mar}$ & REG$_\mathrm{mar}$ & DR$_\mathrm{mar}$ && IPW & REG & DR \\
\midrule
TT & 0.950 & 0.955 & 0.948 & 0.000 & 0.000 & 0.013 && 0.963 & 0.961 & 0.958 \\
FT & 0.000 & 0.939 & 0.949 & 0.000 & 0.000 & 0.000 && 0.040 & 0.960 & 0.970 \\
TF & 0.950 & 0.166 & 0.955 & 0.010 & 0.000 & 0.000 && 0.955 & 0.240 & 0.965 \\
FF & 0.226 & 0.584 & 0.342 & 0.000 & 0.000 & 0.000 && 0.290 & 0.545 & 0.472 \\
\bottomrule
\end{tabular}
\end{table}

In addition to model misspecification, we also evaluate sensitivity of the proposed methods against violation of Assumption~\ref{assump1}.
The difference between the log odds ratios $\Gamma_1$ and $\Gamma_2$ in two calls is used as the sensitivity parameter to capture the degree of departure from the assumption.
The proposed estimators exhibit small bias when the sensitivity parameter varies within a moderate range but it could be large if the stableness assumption is severely violated.
See Section~\ref{ssec:simu-sens} of the Supplementary Material for details.
Therefore, to obtain reliable inference in practice, we suggest using several different working models and recommend the sensitivity analysis for assessing the robustness of inference.

\section{Real data analysis}\label{sec:data}

\subsection{Dataset}
We apply the proposed methods to the NRFU study to estimate the voter turnout in the 2020 U.S. presidential election.
The survey data are publicly available online \citep{nrfu2021}.
After excluding the undeliverable, deceased and removed cases, our analysis utilizes data from 3,892 samples who received questionnaire containing both political content and information on the demographic covariates.
This dataset remains a representative sample of the total population.
The outcome of interest is whether the person voted or not ($Y=1$ or $0$) in the 2020 presidential election.
The respondents include 1,558 people who answered ``I am sure I voted'', 228 who answered ``I am sure I did not vote'', and 21 who answered ``I am not completely sure''.
The contact phase includes two classes of invitation and among the 3,892 samples we analyze, 1,312 responded to the first class invitation questionnaire, 495 responded later to the second class, and 2,085 never responded.

We consider demographic covariates including age, gender, race, ethnicity, and educational attainment,
which are also missing for nonrespondents.
These covariates potentially influence the outcome and the response, and should be controlled for reducing bias and estimation error.
Race is encoded in two categories (white and nonwhite), ethnicity in two categories (Hispanic and Non-Hispanic), age in three categories (18--29, 30--59, and 60+), and educational attainment in three categories (high school or less, some college, and Bachelor's degree or above).
The distribution of these variables is obtained from the U.S. Census data \citep{census2021},
see Section~\ref{ssec:demo-dist} of the Supplementary Material for details.
In addition, there are some fully observed factorial design variables in the NRFU study, including the advance postcard (\textit{m1sent}, sent or not), the questionnaire version (\textit{version}, political content on page 2 or on page 1), study title (\textit{title}, short or long), and prepaid incentive presentation (\textit{incvis}, visible or not);
these covariates are randomized and independent of the demographic covariates, and their distribution is known by design.
Let $X_1=(1, \textit{race}, \textit{ethnicity}, \textit{gender}, \textit{age2}, \textit{age3}, \textit{edu2}, \textit{edu3})^\T$
and $X_2=(\textit{m1sent},\textit{version}, \textit{title}, \textit{incvis})^\T$ denote the demographic and design covariates, respectively, and let $X=(X_1^\T,X_2^\T)^\T$,
where \textit{age2}, \textit{age3}, \textit{edu2} and \textit{edu3} are dummy variables for age and educational attainment categories.
For the 21 respondents who answered ``I am not completely sure'',
they are substantially different from nonrespondents and closer to the respondents.
A reasonable approach is to impute their outcomes using their demographic characteristics.
We first fit a logistic regression model of $Y$ on $X$ among respondents except these 21 units,
and then combine this model and their covariates to impute their voting outcome.

\subsection{Estimation of voter turnout}\label{ssec:5.2}

The primary estimand of interest is the voter turnout rate, that is, $\theta =E(Y)$.
For estimation of $\theta$, we specify working models $\pi_1(x,y;\alpha_1,\gamma)=\operatorname{expit}(\alpha_1^\T x+\gamma y)$, $\pi_2(x,y;\alpha_2,\gamma)=\operatorname{expit}(\alpha_2^\T x+\gamma y)$ and $f_2(y=1\mid x;\beta)=\operatorname{expit}(\beta^\T x)$,
and implement the proposed IPW, REG, and DR estimators.
In addition, we are also interested in $\gamma$ that characterizes the resistance to respond due to the outcome,
and $\alpha_1$ and $\alpha_2$ that encode the association between covariates and the response propensity in the first and second calls, respectively.

For comparison, we also apply standard estimation methods, including the complete-case (CC) estimator, which is the outcome mean of complete cases without adjustment for covariates;
the MAR estimator that adjusts for covariates by assuming missingness at random (MAR),
obtained by letting $\gamma=0$ in the IPW estimating equations \eqref{eq:3}, \eqref{eq:4} and \eqref{eq:6};
the continuum of resistance (COR) estimator,
obtained by substituting the mean of nonrespondents with the mean of respondents in the last call;
the $X$-adjusted COR estimator ($\text{COR}_\text{x}$),
which approximates the conditional outcome distribution of nonrespondents with that of respondents in the last call;
and the traditional multiple imputation (MI) estimator with 50 independent imputations, where the callback data are incorporated as a fully observed covariate.
We also implement two SOR estimators without controlling covariates, including a DR estimator (DR$_0$) and an estimator based on the parameter-counting approach (PC) described in Section~\ref{ssec:s1.2} of the Supplementary Material.

Table~\ref{tbl:4} summarizes the estimation results.
The CC estimate departs far from the VEP voter turnout (0.664), which indicates a large deviation in the turnout rates between the total respondents and nonrespondents.
The MAR and MI estimates are also severely biased from the VEP turnout,
suggesting that solely adjusting for covariates is insufficient to correct the nonresponse bias, even when the callback data are included as a covariate.
The COR and $\text{COR}_\text{x}$ estimates attenuate slightly toward the VEP turnout, but still depart far from the latter,
suggesting a large deviation in the turnout rates between the respondents in the second call and nonrespondents.
Nonetheless, after adjustment with the SOR model, the IPW point estimate of the voter turnout is 0.665 with 95\% confidence interval (0.548, 0.781),
and the DR estimate is close to the IPW estimate.
The REG estimator has slightly larger bias and variance than the IPW and DR estimators, which is likely due to misspecification of the outcome model in this data example.
The IPW, DR and REG estimates of the odds ratio parameter $\gamma$ are positive with 95\% C.I.s far away from zero.
This is evidence for significant nonignorability of the nonresponse.
The odds ratio estimates reveal that people who are less likely to vote in the election are also less likely to respond to the election survey or more difficult to contact.
The proposed DR, IPW and REG estimates are close to the official report of VEP voter turnout, which significantly reduce the nonresponse bias.
In this data example, the SOR model is able to capture this nonignorable missingness mechanism, but the other methods are agnostic to this mechanism and thus fail to correct the nonresponse bias.
Without controlling covariates, DR$_0$ and PC estimates are also effective for correcting the nonresponse bias,
but in general, they are not as good as those with covariates controlled.
In practice, it is crucial to control relevant covariates to achieve the best performance in reducing nonresponse bias and estimation error.

\begin{figure}[!ht]
\centering
\includegraphics[width=0.9\textwidth]{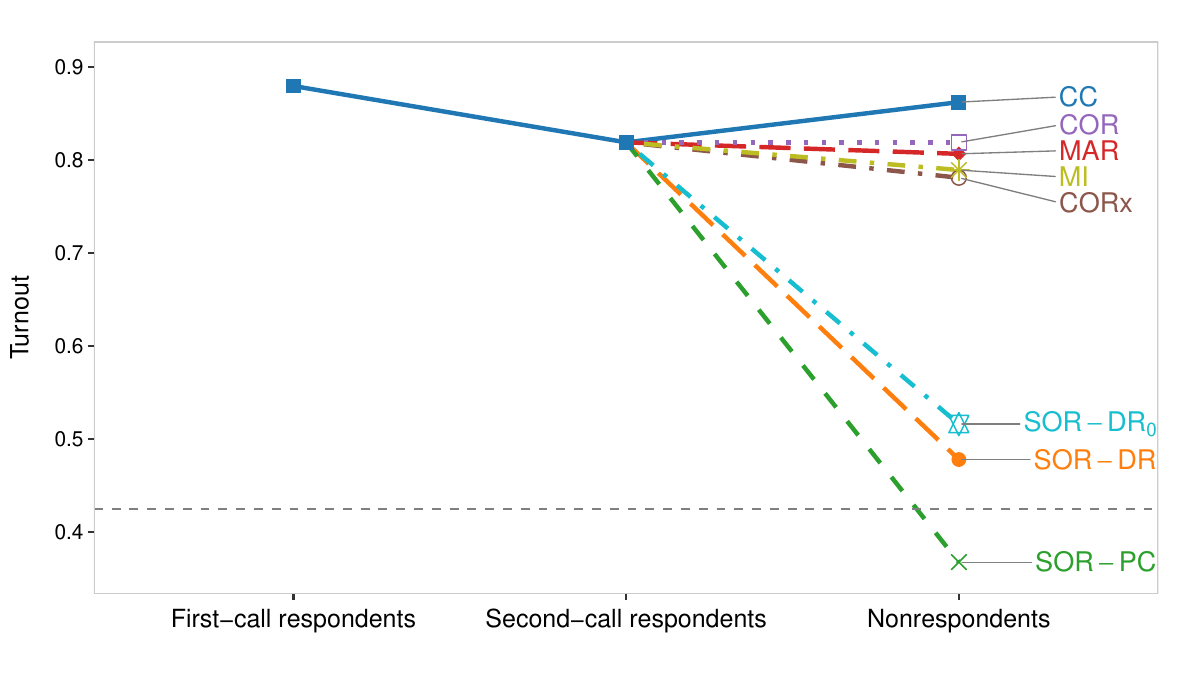}
\caption{Turnout estimation at each contact stage by different methods. The dashed horizontal line marks the turnout of nonrespondents inferred from the VEP turnout.}
\label{fig:3}
\end{figure}

Figure~\ref{fig:3} further illustrates how the turnout rates alter across the contact stages.
The turnout rate for nonrespondents is estimated with CC, MAR, COR, $\text{COR}_\text{x}$, MI and the proposed SOR methods.
The true turnout rate in nonrespondents can be calculated from the official VEP turnout rate,
which shows a dramatic decline as the response unwillingness or contact difficulty increases.
The SOR approach can account for such a decline trend in nonrespondents as it can capture the association between the willingness to vote in the election and the willingness to respond in the survey.
However, the CC and COR methods assume that the turnout rate in the nonrespondents is close to either the total respondents or the latest respondents.
As a result, these methods cannot detect such a decline trend in nonrespondents and therefore lead to considerable overestimation.

\begin{figure}[!ht]
\graphicspath{{figures/Figure_4}}
\includegraphics[scale=0.5]{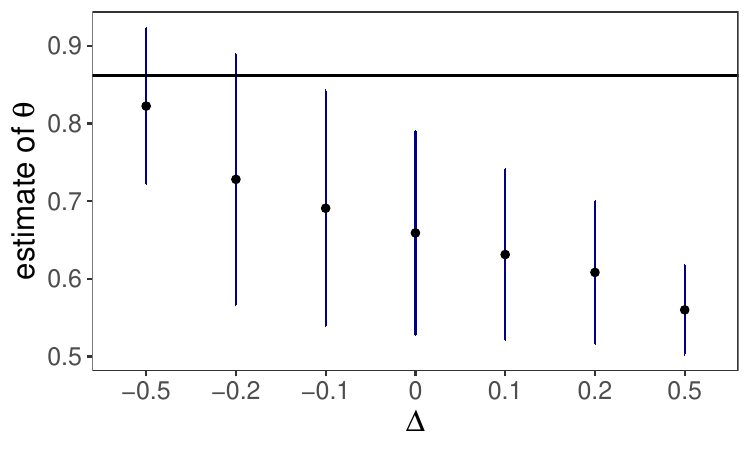}
\hfil
\includegraphics[scale=0.5]{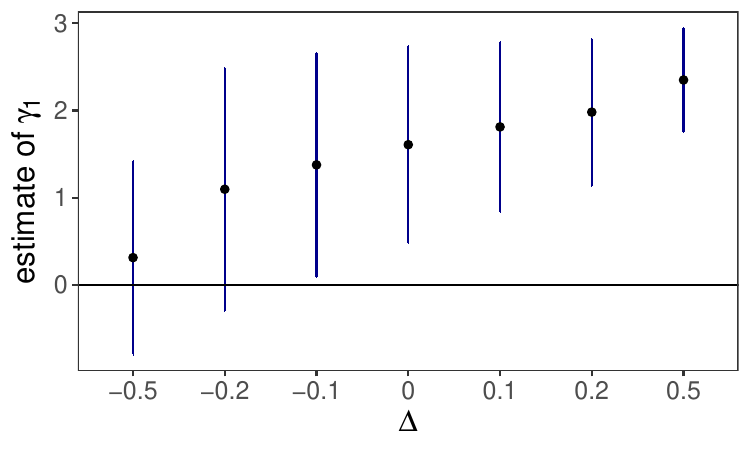}
\caption{Sensitivity analysis of the DR estimation at different values of $\Delta$. The horizontal lines mark the complete-case (CC) sample mean and zero, respectively. Blue bars represent 95\% confidence intervals.}
\label{fig:4}
\end{figure}

The stableness assumption, while plausible when a unit's interest in politics and the local political climate are relatively stable across two adjacent calls, may not hold perfectly in practice.
To assess the robustness of the above results against potential violations of the stableness assumption, we conduct a sensitivity analysis where we allow the log odds ratio parameter to differ between the first two calls by a specified amount $\Delta$.
We use working models $\pi_1 = \expit (\alpha_1^\T X + \gamma_1Y)$ and $\pi_2 = \expit \left\{\alpha_2^\T X +(\gamma_1+\Delta)Y\right\}$,
and use $\Delta$ as the sensitivity parameter,
which takes value in $(-0.5,-0.2,-0.1,0,0.1,0.2,0.5)$.
Figure~\ref{fig:4} summarizes the DR estimates of the turnout and odds ratio parameter $(\theta,\gamma_1)$ under different values of $\Delta$.
When the sensitivity parameter varies within a moderate range, the DR estimates do not change dramatically.
In particular, the estimates of $\theta$ remain lower than the complete-case (CC) sample mean,
and the estimates of $\gamma_1$ remain positive.
Such results reinforce our finding that people who are less likely to vote in the election are also less likely to respond to the election survey.

\begin{table}[!ht]
\caption{Point estimates and 95\% confidence intervals (C.I.) for the voter turnout and the odds ratio parameter}
\label{tbl:4}
\centering
\begin{tabular}{l l l l l}
\toprule
& \multicolumn{2}{c}{\centering \textbf{Turnout rate}} & \multicolumn{2}{c}{\centering \textbf{Odds ratio parameter $\gamma$}} \\
\cmidrule{2-5}
{Method} & {Estimate} & {95\% C.I.} & {Estimate} & 95\% C.I.\\
\midrule
\textit{CC} & 0.862 & (0.842, 0.881) & --- & ---\\
\textit{MAR} & 0.806 & (0.779, 0.833) & --- & ---\\
\textit{COR} & 0.842 & (0.832, 0.853) & --- & ---\\
\textit{$\text{COR}_{\text{x}}$} & 0.801 & (0.763, 0.839) & --- & ---\\
\textit{MI} & 0.829 & (0.752, 0.906) & --- & ---\\
SOR methods: &&&&\\
\textit{PC} & 0.638 & (0.507, 0.769) & --- & ---\\
\textit{DR$_0$} & 0.729 & (0.527, 0.932) & 0.114 & (-1.641, 1.869) \\
\textit{IPW} & 0.665 & (0.548, 0.781) & 1.549 & (0.337, 2.761) \\
\textit{REG} & 0.611 & (0.470, 0.753) & 2.339 & (0.956, 3.722) \\
\textit{DR} & 0.659 & (0.528, 0.790) & 1.608 & (0.482, 2.733)\\
\bottomrule
\end{tabular}
\end{table}

\subsection{The association between covariates and nonresponse as well as turnout}\label{ssec:5.3}
The SOR approach also allows us to analyze the association between covariates and the response propensity at each contact stage.
Table~\ref{tbl:5} reports the point estimates and 95\% confidence intervals for covariates coefficients in the propensity score models.
From Table~\ref{tbl:5},
the advance postcard, questionnaire version, title length, and incentive presentation do not show discernible effects on the response propensity in the first or second contacts.
Mailed advance postcards may not promote response, and it could be skipped in future surveys to reduce cost;
this has also been discovered by \citet{debell2022experimental} in the NRFU study.
In terms of demographic characteristics, gender, age and ethnicity are significantly associated with response rates.
In particular, female and senior people are more likely to respond in each contact,
suggesting female and senior people are more active in political surveys or easier to reach by mail.
Besides, the association between gender and the response propensity increases with repeated contacts,
suggesting that repeated contacts are more useful to encourage females to respond compared to males.
In addition, Hispanics are less likely to respond, especially in the second contact.
These findings offer valuable insights for the design and promotion strategies in election surveys, assisting researchers in developing more effective and feasible methods to attract diverse demographic groups and encourage their participation in surveys.

\begin{table}[!ht]
\caption{Point estimates and 95\% confidence intervals (C.I.) for covariates coefficients in propensity score models}
\label{tbl:5}
\centering
\begin{tabular}{l l l l l}
\toprule
& \multicolumn{2}{c}{\centering \textbf{First contact}} & \multicolumn{2}{c}{\centering \textbf{Second contact}} \\
\cmidrule{2-5}
{Covariate} & {Estimate} & {95\% C.I.} & {Estimate} & 95\% C.I.\\
\midrule
\textit{intercept} & -2.225 & (-2.817, -1.633) & -2.540 & (-3.754, -1.326) \\
\textit{m1sent}: advance postcard & -0.181 & (-0.478, 0.116) & -0.207 & (-0.755, 0.342) \\
\textit{version}: on page 1 & 0.204 & (-0.089, 0.497) & 0.205 & (-0.266, 0.677) \\
\textit{title}: long & -0.054 & (-0.347, 0.240) & -0.257 & (-0.807, 0.294)\\
\textit{incvis}: visible & 0.107 & (-0.189, 0.403) & 0.180 & (-0.342, 0.702)\\
\textit{race}: nonwhite & -0.166 & (-0.559, 0.227) & 0.359 & (-0.226, 0.944)\\
\textit{ethnicity}: Hispanic & -0.429 & (-0.913, 0.056) & -0.764 & (-1.448, -0.081)\\
\textit{gender}: male & -0.434 & (-0.729, -0.139) & -0.689 & (-1.296, -0.082) \\
\textit{age2}: 30--59 & 0.596 & (0.172, 1.021) & 0.574 & (-0.043, 1.190)\\
\textit{age3}: 60+ & 1.352 & (0.861, 1.842) & 1.698 & (1.031, 2.364) \\
\textit{edu2}: some college & 0.220 & (-0.327, 0.766) & 0.253 & (-0.606, 1.111)\\
\textit{edu3}: Bachelor's degree or above & 0.326 & (-0.370, 1.022) & 0.383 & (-0.672, 1.438)\\
\bottomrule
\end{tabular}
\end{table}

We also investigate the association between demographic covariates and voting after the adjustment of nonresponse, which we call the voting-demographic model.
The parameter of interest now is the regression coefficients
in the logistic regression of $Y$ on $X_1$,
and the corresponding estimating function defining the parameter is $m(x, y ; \theta)=x_1\{y-\operatorname{expit}(x_1^\T\theta)\}$.
We summarize in Table~\ref{tbl:6} the estimates of the coefficients obtained with the proposed IPW method.
The results suggest that age and educational attainment are significantly associated with voting,
and senior and better-educated people are more likely to vote.
Hispanics have significantly lower turnout at the 10\% level (90\% C.I. (-0.826, -0.066)).
The results do not show strong association of gender or race with voting.
The estimate of intercept in the respondents (-0.011, 95\% C.I. (-0.527, 0.505)) is much larger than that after adjusting for nonresponse (-1.095, 95\% C.I. (-1.941, -0.249)),
suggesting that the respondents-based logistic regression may severely overestimate turnout.
Although, additional research is required to further investigate the reliability and robustness of these findings in a broader context.

\begin{table}[!ht]
\caption{Point estimates and 95\% confidence intervals (C.I.) for covariates coefficients in the voting-demographic model}
\label{tbl:6}
\centering
\begin{tabular}{l l l l l}
\toprule
& \multicolumn{2}{c}{\centering \textbf{CC estimates}} & \multicolumn{2}{c}{\centering \textbf{SOR estimates}} \\
\cmidrule{2-5}
{Covariate} & {Estimate} & {95\% C.I.} & {Estimate} & 95\% C.I.\\
\midrule
\textit{intercept} & -0.011 & (-0.527, 0.505) & -1.095 & (-1.941, -0.249) \\
\textit{race}: nonwhite & 0.091 & (-0.328, 0.509) & 0.119 & (-0.317, 0.554)\\
\textit{ethnicity}: Hispanic & -0.198 & (-0.693, 0.297) & -0.446 & (-0.899, 0.007)\\
\textit{gender}: male & 0.024 & (-0.316, 0.364) & -0.071 & (-0.527, 0.384) \\
\textit{age2}: 30--59 & 0.491 & (0.015, 0.967) & 0.547 & (0.059, 1.034)\\
\textit{age3}: 60+ & 1.295 & (0.781, 1.810) & 1.710 & (1.080, 2.340) \\
\textit{edu2}: some college & 1.327 & (0.934, 1.721) & 1.474 & (1.045, 1.904)\\
\textit{edu3}: Bachelor's degree or above & 2.243 & (1.813, 2.673) & 2.406 & (1.909, 2.902)\\
\bottomrule
\end{tabular}
\begin{tablenotes}[hang]
\item[]Note: The column ``CC estimates'' corresponds to estimates obtained by logistic regression with complete cases.
The column ``SOR estimates'' corresponds to estimates by the proposed SOR-IPW method.
\end{tablenotes}
\end{table}

\subsection{Heterogeneity of response}\label{ssec:heterogeneity}

We conduct additional analysis to investigate heterogeneous effects of the design covariates on the response and heterogeneous nonignorability across demographic groups.
Specifically, we examine the interaction of the visible cash incentive with gender and age, and the interaction of voting and gender in the propensity score models.
Table~\ref{tbl:s.ita} of the Supplementary Material shows a statistically significant interaction effect between visible cash incentive and gender on first-call response (point estimate 0.676, 95\% C.I. (0.108, 1.244)), suggesting that visible cash incentives increase response rates more effectively among men than women,
and thus, proper cash incentives could partially compensate for overrepresentation in gender-imbalanced samples.
This finding aligns with prior evidence that monetary incentives tend to motivate men more strongly than women in task performance \citep[e.g.,][]{czap2018comparing,spreckelmeyer2009anticipation}.

\subsection{Estimation of Trump's popular vote}\label{ssec:trump-supp}
The underestimation of Trump support is also noteworthy in the ANES.
We apply the proposed SOR approach to estimate Trump's popular vote in the 2020 presidential election,
defining $Y=1$ for voting Trump, $Y=0$ for voting others and $Y=-1$ for not voting.
The parameter of interest is Trump's popular vote, that is, $\theta =P(Y=1)/\{P(Y=1)+P(Y=0)\}$.
As shown in Table~\ref{tbl:7}, standard estimates---CC (0.416), MAR (0.421), COR (0.417) and COR$_\text{x}$ (0.411)---all fall approximately 0.05 below the official report (0.468).
The proposed DR estimate (0.428, 95\% C.I. (0.363, 0.493)) is close to these, and
the estimate of the odds ratio parameter $\gamma$ (-0.399, 95\% C.I. (-1.009, 0.210)) associated with voting Trump versus voting others provides no strong evidence of nonignorable nonresponse.
Previously, \citet{bailey2025countering}'s analysis of Trump approval in a 2019 Ipsos Knowledge Panel survey suggests that while nonignorable nonresponse is not detected when the data are not stratified by region and party, it becomes significant once these variables are incorporated.
This reinforces the heterogeneous nature of nonignorability by subgroups, which cannot be fully examined in the current study because we have no access to the distribution of such covariates in the census data.
We emphasize incorporating more politically directional covariates in future studies and refer to \citet{bailey2025countering} for a practical strategy and insightful discussion.

\begin{table}[!ht]
\caption{Point estimates and 95\% confidence intervals (C.I.) for Trump's popular vote and the odds ratio parameter}
\label{tbl:7}
\centering
\begin{tabular}{l l l l l}
\toprule
& \multicolumn{2}{c}{\centering \textbf{Trump's popular vote}} & \multicolumn{2}{c}{\centering \textbf{Odds ratio parameter $\gamma$}} \\
\cmidrule{2-5}
{Method} & {Estimate} & {95\% C.I.} & {Estimate} & 95\% C.I.\\
\midrule
\textit{CC} & 0.416 & (0.388, 0.443) & --- & ---\\
\textit{MAR} & 0.421 & (0.390, 0.451) & --- & ---\\
\textit{COR} & 0.417 & (0.402, 0.431) & --- & ---\\
\textit{$\text{COR}_{\text{x}}$} & 0.411 & (0.372, 0.449) & --- & ---\\
\textit{DR} & 0.428 & (0.363, 0.493) & -0.399 & (-1.009, 0.210)\\
\bottomrule
\end{tabular}
\end{table}

\section{Discussion}

Nonresponse remains a crucial issue in election surveys. 
This article demonstrates how the SOR model, with callback and census data, can adjust for nonignorable missingness in both outcomes and covariates. 
Our approach captures declining likelihood of voting among more reluctant respondents and produces turnout estimates close to official statistics. 
The SOR assumption is most plausible when political interest and local environment remain stable across adjacent call attempts, though it may break down over longer periods. 
Its justification requires case-specific domain knowledge, and sensitivity analysis is essential to evaluate robustness against possible violations.
Our work complements the RRI framework of \citet{bailey2024polling,bailey2025countering}. 
Whereas RRI relies on randomized survey features, SOR leverages callback data without randomization. 
Although the NRFU study encompasses an RRI design, the RRI approach is not applicable here due to weak association between randomized design features and response as shown in Section~\ref{ssec:rri} of the Supplementary Material. 
In contrast, the two-stage callback design made SOR particularly suitable here. 
Future studies should combine both approaches when applicable to exploit their complementary strengths.
Finally, while we focus on nonresponse bias, overreporting also threatens survey validity. 
This issue may be less severe in the NRFU study, which uses post-election mail surveys, but remains a concern for pre-election settings. 
Addressing both overreporting and nonresponse is critical for improving political survey accuracy.

\paragraph{Acknowledgments}
We are grateful to valuable comments from the editor and four anonymous reviewers.
This work is partially supported by the National Natural Science Foundation of China (42450214) and the National Key R\&D Program of China (2022YFA1008100).

\paragraph{Data Availability Statement}
Replication code for this article is available at \citet{li2026replication} at https://doi.org/10.7910/DVN/TNABTF.

\paragraph{Supplementary Material}
The Supplementary Material contains further illustration for the stableness of resistance, estimation methods with multiple callbacks, proof of theorems and unbiasedness of estimating equations, and additional simulations and real data analysis results.

\putbib
\end{bibunit}

\newpage

\appendix

\setcounter{assumption}{0}
\setcounter{lemma}{0}
\setcounter{table}{0}
\setcounter{figure}{0}
\setcounter{theorem}{0}
\setcounter{equation}{0}

 {\centering \section*{Supplementary  Material}}

\renewcommand {\thesection} {S\arabic{section}}
\renewcommand {\theexample} {S.\arabic{example}}
\renewcommand {\theassumption} {S.\arabic{assumption}}
\renewcommand {\thetheorem} {S.\arabic{theorem}}
\renewcommand {\theequation} {S.\arabic{equation}}
\renewcommand {\thetable} {S.\arabic{table}}
\renewcommand {\thefigure} {S.\arabic{figure}}

This supplement contains further illustration for the stableness of resistance, estimation methods with multiple callbacks, proof of theorems and unbiasedness of estimating equations, and additional simulations and real data analysis results.

\begin{bibunit}

\section{Further illustration for the stableness of resistance}

\subsection{Illustration for the stableness of resistance via a generative model}\label{ssec:s1.1}

To further  illustrate the intuition behind Assumption~\ref{assump1} in election surveys,
we provide a generative model about how a unit makes choice to vote and decides to respond in the election survey, 
which is motivated from  the discrete choice theory \citep{mcfadden2001economic}.
Suppose
\begin{equation}\label{eq:2}
\begin{aligned}
&U_0=\beta_0+\beta_1^\T X+\beta_2C+\varepsilon_0, \quad Y=I(U_0>0),\\
    &U_1 =\alpha_{10}+\alpha_{11}^\T X+\gamma_1C+\varepsilon_1,\quad U_2 =\alpha_{20}+\alpha_{21}^\T X+\gamma_2C+\varepsilon_2,\\
    & R_1=I(U_1>0),\quad R_2=R_1+(1-R_1)\times I(U_2>0),\\
    &(\varepsilon_0,\varepsilon_1,\varepsilon_2)\perp (X,C),\quad \varepsilon_0\perp \varepsilon_1 \perp \varepsilon_2,
\end{aligned}
\end{equation}
where the outcome $Y$  is a binary variable about whether a person voted or not, 
$X$ is a vector of observed covariates, 
and $C$ denotes an unmeasured factor. 
The unmeasured factor $C$ is correlated with both the outcome and missingness, 
which is the source for nonignorable nonresponse.
For example,   $C$ could include a unit's interest in politics or local political climate \citep{marsh2002electoral,camatarri2023always},
which is correlated with  the voting behavior and also an important source for   nonresponse to election surveys.
Variable $U_0$ stands for  the utility or net benefit that a unit obtains from voting as opposed to not voting, 
and $U_1$ and $U_2$ are the utilities that a unit decides to respond or not at the first and second calls, respectively. 
The error terms $(\varepsilon_0,\varepsilon_1,\varepsilon_2)$ are mutually independent  and independent of $X$ and $C$, 
following  logistic distributions.
The generating process of $Y,R_1$ and $R_2$ reveals that 
a unit will vote if utility $U_0>0$ and will respond in the first call (i.e. $R_1=1$) if utility $U_1>0$.
Those who did not respond in the first call will respond at the second call (i.e. $R_1=0,R_2=1$) if utility $U_2>0$.
The parameters $\beta_2,\gamma_1,\gamma_2$ capture how the unmeasured factor $C$ affect the voting behavior and the response propensity in different calls.
Under this generative model, the propensity scores in  the two calls can be well approximated with logistic models.
In particular, these two logistic models have approximately equal odds ratio parameters if $\gamma_1=\gamma_2$.
Therefore, Assumption~\ref{assump1} about the propensity scores
essentially approximates that $\gamma_1=\gamma_2$ in the generative model, and thus, we can justify   Assumption~\ref{assump1} by assessing whether the influence of the unmeasured factor on the utilities of responding  remains stable  in the first two calls. 

\begin{figure}[!ht]
\graphicspath{{figures/Figure_S1}}
 \begin{minipage}{0.45\linewidth}
 	\centerline{\includegraphics[width=\textwidth]{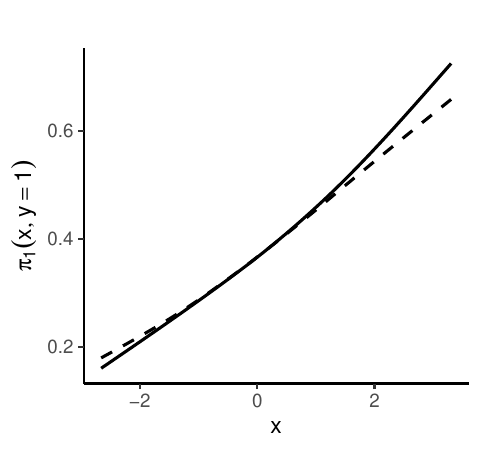}}
 	\centerline{\includegraphics[width=\textwidth]{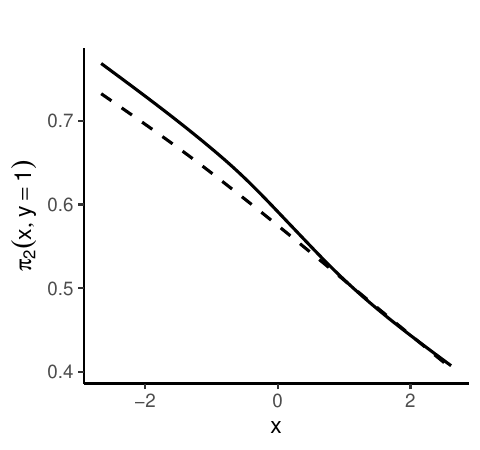}}
 \end{minipage}
 \begin{minipage}{0.45\linewidth}
	\centerline{\includegraphics[width=\textwidth]{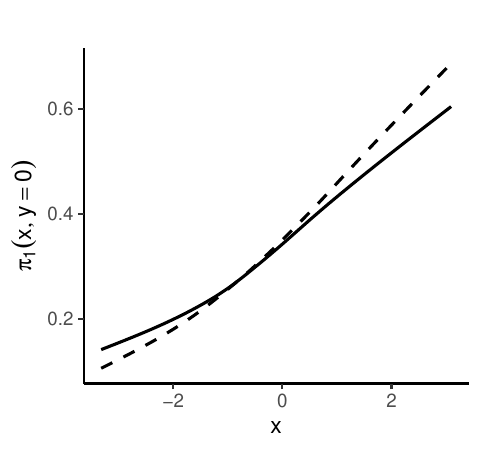}}
	\centerline{\includegraphics[width=\textwidth]{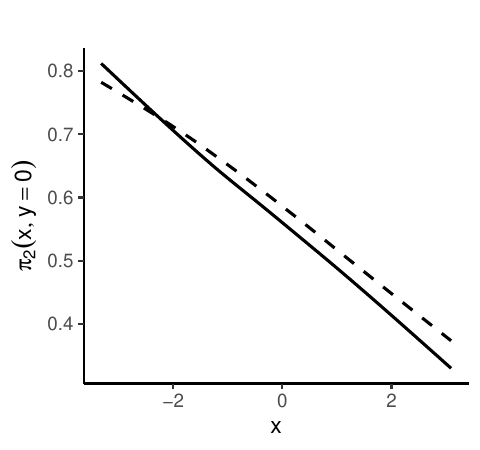}}
\end{minipage}
\caption{The response propensity scores of the first two calls and the fitted logistic models.} \label{fig:s.1}
\end{figure}

\begin{figure}[!ht]
\graphicspath{{figures/Figure_S2}}
\begin{minipage}{0.45\linewidth}
		\centerline{\includegraphics[width=\textwidth]{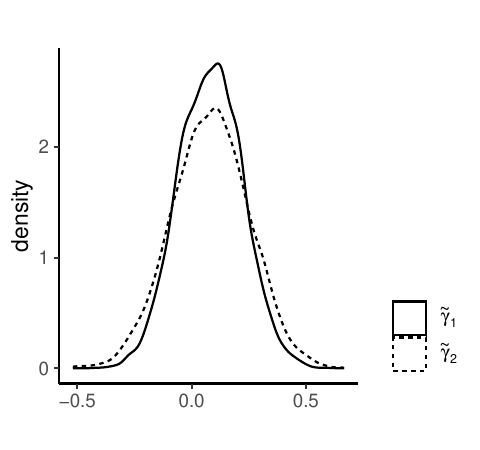}}
	\end{minipage}
	\begin{minipage}{0.45\linewidth}
		\centerline{\includegraphics[width=\textwidth]{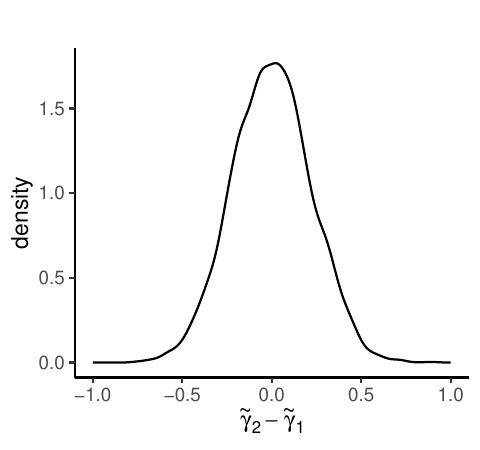}}
	\end{minipage}
\caption{The density plots of two odds ratio parameters in the fitted propensity score models and their differences.} \label{fig:s.2}
\end{figure}

Here are some detailed numerical illustrations about  the connection between Assumption~\ref{assump1} and the generative model \eqref{eq:2}.
We generate data from model \eqref{eq:2} with 
\begin{eqnarray*}
    &&(X,C)^\T\sim N(0,I_2),\quad (\beta_0,\beta_1,\beta_2)=(0.2,0.35,0.3),\\
    &&(\alpha_{10},\alpha_{11},\alpha_{20},\alpha_{21})=(-0.6,0.4,0.35,-0.3), \quad \gamma_1=\gamma_2=0.3.
\end{eqnarray*}
Note that in this case the influence of the unmeasured factor on the willingness to respond are the same in these two calls.
With simulated data of sample size 1,000,
Figure~\ref{fig:s.1} shows the true response propensity (solid line) and  the approximation of logistic regression curve (dashed line) for  each call as a function of $X$ among voters ($Y=1$) and nonvoters ($Y=0$), respectively.
In each case, the true propensity score  and the fitted logistic regression curve are close.
Then we fit logistic regression of $R_1$ on $(X,Y)$ to obtain the odds ratio estimate $\tilde{\gamma}_1$, 
and fit logistic regression of $R_2$ on $(X,Y)$ among the nonrespondents to  the first call $(R_1=0)$ to obtain the odds ratio estimate $\tilde{\gamma}_2$.
Figure~\ref{fig:s.2} shows the density plot of the regression coefficients $\tilde{\gamma}_1$ and $\tilde{\gamma}_2$, 
and of their differences based on 5,000 simulation replicates.
The differences of these two coefficients are concentrated around zero,
which suggests that the  odds ratio parameters in logistic regression models are approximately the same in these two calls.

\subsection{A parameter-counting interpretation of identification in the case of a binary outcome}\label{ssec:s1.2}

For a binary outcome, we provide some intuition on the role of Assumption~\ref{assump1} using a parameter-counting argument.
Conditioning on each level of the covariate $X = x$ (we suppress the covariate for simplicity), there are $6$ unknown parameters for the joint distribution $f(Y,R_1,R_2)$ because $R_2\geq R_1$ by definition,
	 \begin{equation*}
	 	\begin{aligned}
	 		p_1 = f(Y = 0, R_1 = 0, R_2 = 0), \quad p_2 = f(Y = 1, R_1 = 0, R_2 = 0), \quad p_3 = f(Y = 0, R_1 = 0, R_2 = 1), \\
	 		p_4 = f(Y = 1, R_1 = 0, R_2 = 1), \quad p_5 = f(Y = 0, R_1 = 1, R_2 = 1), \quad p_6 = f(Y = 1, R_1 = 1, R_2 = 1).
	 	\end{aligned}
	 \end{equation*}
Four parameters $p_3, p_4, p_5, p_6$ are identifiable from the observed data, with a natural constraint that $p_1 + p_2 + p_3 + p_4 + p_5 + p_6 = 1$.
In total, there are $5$ equations, which do not suffice to identify $6$ unknown parameters without imposing additional assumption.
 
The log odds ratio functions for the response propensity in the first two calls can be written as 
 \begin{eqnarray*}
 	\Gamma_1(Y=1) = \log\frac{p_6(p_1+p_3)}{p_5(p_2+p_4)}, \quad \Gamma_2(Y=1) = \log\frac{p_4p_1}{p_2p_3}.
 \end{eqnarray*}
The stableness of resistance assumption (Assumption~\ref{assump1}) requires that $\Gamma_1 = \Gamma_2$, i.e.,
 \begin{equation}\label{eq:s.1}
 	\frac{p_6(p_1 + p_3)}{p_5(p_2 + p_4)} = \frac{p_4p_1}{p_2p_3}.
 \end{equation}
The additional restriction \eqref{eq:s.1} together with the previous $5$ equations are sufficient to identify $6$ unknown parameters.
The solution to these equations is unique and $p_1,p_2$ can be written as functions of $(p_3,p_4,p_5,p_6)$.

It is possible to consider  other constraints for identification, such as stableness on the risk ratio scale.
The risk ratio functions for the response propensity in the first two calls are 
\begin{eqnarray*}
	\Lambda_1(X,Y) = \log \frac{f(R_1 = 1\mid X,Y)}{f(R_1 = 1\mid X,Y=0)}, \quad \Lambda_2(X,Y) = \log \frac{f(R_2 = 1\mid R_1=0, X,Y)}{f(R_2 = 1\mid R_1=0,X,Y=0)}.
\end{eqnarray*}
For  a binary outcome with the covariate surpressed,
 \begin{eqnarray*}
	\Lambda_1(Y=1) = \log\frac{p_6(p_1+p_3+p_5)}{p_5(p_2+p_4+p_6)}, \quad \Lambda_2(Y=1) = \log\frac{p_4(p_1+p_3)}{p_3(p_2+p_4)}.
\end{eqnarray*}
The stableness of resistance   on risk ratio scale, which states $\Lambda_1 = \Lambda_2$, is equivalent to
\begin{equation}\label{eq:s.2}
	 \frac{p_6(p_1 + p_3 + p_5)}{p_5(p_2 + p_4 + p_6)} = \frac{p_4(p_1 + p_3)}{p_3(p_2 + p_4)}.
\end{equation}
Analogous to equation \eqref{eq:s.1},  equation \eqref{eq:s.2} imposes an additional constraint on the parameters, and
identification of the unknown parameters  could also be achievable.
By introducing other constraints on the unknown parameters, we may obtain different sufficient conditions for identification in the case of a binary outcome.
However, the interpretation and justification of different constraints need to be further investigated.
We recommend the stableness of resistance assumption because it is theoretical guaranteed and interpretable in election surveys.

\section{Estimation methods with multiple callbacks}\label{sec:multi-callback}
Suppose we have $K(\geq3)$ calls and let $R_k$ denote the response state in the $k$th call for $k=1,\ldots, K$.
We illustrate how to construct IPW, REG and DR estimators that make full use of data from all calls.

For the IPW estimation, to incorporate   observations on $(X,Y)$ after the second call we specify a working model for  $\pi_K(x,y;\alpha_K,\gamma_K) =f(R_K=1\mid R_2=0,X=x,Y=y)$ in addition to $\pi_1(\alpha_1,\gamma),\pi_2(\alpha_2,\gamma)$,
where $\alpha_K$ and $\gamma_K$ are the parameters of the baseline propensity score 
and the odds ratio function of $f(R_K=1\mid R_2=0,X,Y)$, respectively. 
Note that the model for $\pi_K$ is completely independent of $\pi_1,\pi_2$ and does not need  to have the same odds ratio as them. 
We solve \eqref{eq:3}--\eqref{eq:5}  to obtain the estimators $(\hat\alpha_{1,\ipw},\hat\alpha_{2,\ipw},\hat\gamma_{\ipw})$, which remains  the same as the setting with two calls; we solve  \eqref{eq:ipws.1} to obtain the estimators $(\hat\alpha_{K,\ipw},\hat\gamma_{K,\ipw})$.
\begin{eqnarray}
	0&=&\sum_{i=1}^n  w_i\left[\left\{\frac{r_{K,i} - r_{2,i}}{ \pi_K(\hat\alpha_{K,\ipw},\hat\gamma_{K,\ipw})} -  \frac{1 - \hat p_{2,i}}{\hat p_{2,i}}r_{2,i} \right\}\cdot W(x_i,y_i)\right],  \label{eq:ipws.1}
\end{eqnarray}
where $\hat p_{2,i}=\{1-\pi_{1,i}(\hat\alpha_{1,\ipw},\hat\gamma_{\ipw})\}\pi_{2,i}(\hat\alpha_{2,\ipw},\hat\gamma_{\ipw}) + \pi_{1,i}(\hat\alpha_{1,\ipw},\hat\gamma_{\ipw})$  and  $W(x,y)$ is a user-specified vector function of the same dimensional as $(\alpha_K,\gamma_K)$. 
An IPW estimator of $\theta$ using observed data from all callbacks is 
\begin{eqnarray}
	\hat \theta_{K,\ipw} = \sum_{i=1}^n w_i\frac{r_{K,i}y_i}{\hat p_{K,i}} , \label{eq:ipws.2}
\end{eqnarray}
where $\hat p_{K,i}=\hat p_{2,i} + (1-\hat p_{2,i})\hat\pi_{K,i}$ is an estimator of $p_K=f(R_K=1\mid X,Y)$.
Equations \eqref{eq:3}--\eqref{eq:5} are unbiased   for   estimating  $(\alpha_1,\alpha_2,\gamma)$ if $\pi_1(\alpha_1,\gamma),\pi_2(\alpha_2,\gamma)$ are correct
and   \eqref{eq:ipws.1} is   unbiased   for estimating $(\alpha_K,\gamma_K)$ if further $\pi_K(\alpha_K,\gamma_K)$ is correct,
which yield a consistent estimator of $(\alpha_1,\alpha_2,\gamma,\alpha_K,\gamma_K)$.
Therefore, $\hat \theta_{K,\ipw} $ is consistent if $\{\pi_1(\alpha_1,\gamma),\pi_2(\alpha_2,\gamma), \pi_K(\alpha_K,\gamma_K)\}$
are correct.

For  the REG estimation,  we specify working models for  $\pi_1(\alpha_1,\gamma)$, $f_2(y\mid x;\beta),  f_K(y\mid x;\beta_K)=f(Y=y\mid X=x, R_K=1,R_2=0;\beta_K)$, and the log odds ratio function   $\Gamma_K(x,y;\gamma_K)$ of $f(R_K=1\mid R_2=0,X=x,Y=y)$.
We solve \eqref{eq:7}--\eqref{eq:8} to obtain  $(\hat\alpha_{1,\reg}, \hat\gamma_\reg,\hat\beta_{\reg})$,
which remains  the same as the setting with two calls;
we solve  \eqref{eq:regs.1}--\eqref{eq:regs.2} to obtain $(\hat\beta_{K,\reg}, \hat\gamma_{K,\reg})$.
\begin{eqnarray}
    0&=&\sum\limits_{i=1}^n w_i \left\{  (r_{K,i}-r_{2,i}) \cdot \left. \frac{ \partial \log f_K(y_i\mid x_i;\beta_K)}{\partial \beta_K}\right|_{\beta_K=\hat \beta_{K,\reg}} \right\},\label{eq:regs.1} \\
    E_{ f}\{h_{K,W}(X;\hat{\beta}_{K,\reg},\hat{\gamma}_{K,\reg})\}&=&\sum\limits_{i=1}^n w_i\left[  
\begin{aligned}
\left\{\frac{r_{1,i}}{\pi_{1,i}(\hat \alpha_{1,\reg},\hat \gamma_{\reg})} -r_{K,i}\right\} W(x_i,y_i)\\
+  r_{K,i} h_{K,W}(x_i;\hat\beta_{K,\reg},\hat \gamma_{K,\reg})  
\end{aligned}
\right],
\label{eq:regs.2}
\end{eqnarray} 
where $W(x,y)$ is a user-specified function with the same dimension as $\gamma_K$, and $h_{K,W}(x;\beta_K,\gamma_K)=E\{W(X,Y) \mid X=x, R_K=0;\beta_K,\gamma_K\}$ imputes the missing values of $W(X,Y)$.

Then we solve the following  estimating equation to obtain the estimator $\hat\theta_{K,\reg}$, 
\begin{equation}\label{eq:regs.3}
	0  = \sum\limits_{i=1}^n w_i\left[  r_{K,i} \Big\{m(x_i,y_i;\hat\theta_{K,\reg})-h_{K,m}(x_i;\hat\theta_{K,\reg},\hat\beta_{K,\reg},\hat\gamma_{K,\reg}) \Big\}\right]+E_{ f}\Big\{h_{K,m}(X;\hat\theta_{K,\reg},\hat\beta_{K,\reg},\hat\gamma_{K,\reg})\Big\},
\end{equation}
where $h_{K,m}(x;\theta_K,\beta_K,\gamma_K)=E\{m(X,Y;\theta_K) \mid X=x, R_K=0;\beta_K,\gamma_K\}$ imputes the missing values of $m(X,Y;\theta_K)$.

In the presence of multiple callbacks, the doubly robust estimator $\hat \theta_\dr$ described in Section~\ref{ssec:dr}   is agnostic to the observed data after the second call.
We construct a doubly/multiply roust (MR) estimator  that incorporates all observed data on $(X,Y)$, 
although its efficiency is not assessed.
We specify working models    $\pi_1(\alpha_1,\gamma), \pi_2(\alpha_2,\gamma)$, $f_2(\beta)$, $\pi_K(\alpha_K,\gamma_K)$, and $f_K(\beta_K)$;
we solve \eqref{eq:10}--\eqref{eq:13} for $(\hat\alpha_{1,\dr},  \hat\gamma_\dr)$ together with \eqref{eq:regs.1} for $\hat\beta_{K,\dr}$ and \eqref{eq:drs.1} for $(\hat\alpha_{K,\dr}, \hat\gamma_{K,\dr})$,
\begin{equation}
	0= \sum\limits_{i=1}^n w_i\left[ \left\{\frac{r_{K,i} - r_{2,i}}{ \pi_{K,i}(\hat\alpha_{K,\dr},\hat\gamma_{K,\dr})} + r_{2,i} -  \frac{r_{1,i}}{\hat\pi_{1,i}} \right\} \cdot
	\left\{\begin{array}{l}
		V_3(x_i)\\
		U(x_i,y_i) -  h_{K,U}(x_i,y_i;\hat\beta_{K,\dr},\hat\gamma_{K,\dr}) 
	\end{array}
	\right\}\right].  \label{eq:drs.1}
\end{equation}
Then the DR/MR estimator $\hat \theta_{K,\dr}$  is obtained by solving 
\begin{eqnarray}
	0 & =& \sum\limits_{i=1}^n w_i\left[ \left\{ r_{2,i} + \frac{r_{K,i}-r_{2,i}}{\pi_{K,i}(\hat \alpha_{K,\dr},\hat\gamma_{K,\dr})}\right\}\{m(x_i,y_i;\hat\theta_{K,\dr})-h_{K,m}(x_i;\hat\theta_{K,\dr},\hat\beta_{K,\dr},\hat\gamma_{K,\dr})\} \right]\nonumber\\
    &&+E_{ f}\Big\{h_{K,m}(X;\hat\theta_{K,\dr},\hat\beta_{K,\dr},\hat\gamma_{K,\dr})\Big\}, \label{eq:drs.2}
\end{eqnarray}
We have shown in Section~\ref{ssec:dr} that estimators $(\hat\gamma_\dr, \hat\theta_\dr)$ are doubly robust against misspecification of 
$A_2(\alpha_2)$ and $f_2(\beta)$, provided that $\pi_1(\alpha_1,\gamma)$ is correct.
We further have the $2\times 2$-multiply robustness of $\hat\gamma_{K,\dr}$ and $\hat\theta_{K,\dr}$.
\begin{theorem}\label{theorems.1}
	Under the conditions of Theorem~\ref{theorem1}, a stronger  positivity assumption that $c<\pi_k(X,Y)<1-c$, $k=1,2,K$, and   regularity conditions described by \citet[][Theorems 2.6 and 3.4]{newey1994large},  estimators  $(\hat\gamma_{K,\dr},\hat\theta_{K, \dr})$ are  consistent and asymptotically normal provided  that
	$A_1(\alpha_1),\Gamma(\gamma)$, $\Gamma_K(\gamma_K)$, at least one of   $A_2(\alpha_2), f_2(\beta)$,
	and at least one of $A_K(\alpha_K),f_K(\beta_K)$ are correctly specified.
\end{theorem}
See Section~\ref{ssec:s3.4} for the proof of Theorem~\ref{theorems.1}.
It is of  further interest to construct $2^{K-1}$--multiply robust estimators in the sense of \citet{vansteelandt2007estimation} when    propensity scores and outcome distributions for all calls are modeled.
We defer this to future work.

\section{Proof of theorems and unbiasedness of estimating equations}

\subsection{Proof of Theorem~\ref{theorem1}}\label{ssec:s3.1}
First, consider the case that $K=2$.
The observed-data likelihood is
\begin{align*}
    f(\mathcal{O})=f(R_1=1,Y,X)^{R_1}f(R_2=1,R_1=0,Y,X)^{R_2-R_1}f(R_2=0,R_1=0)^{1-R_2}.
\end{align*}
By the data structure, we have
\begin{eqnarray*}
    f(X) 
	& = & f(R_1=0,X)+f(R_1=1,X)\\
	& = & f(R_2=0,R_1=0,X)+f(R_2=1,R_1=0,X) +f(R_1=1,X).
\end{eqnarray*}
Because $f(R_2=1,R_1=0,X)$ and $f(R_1=1,X)$ are available from the observed data and the sampling weights, $f(R_2=0,R_1=0,X)$ and $f(R_1=0,X)$ can be obtained when $f(X)$ is known.

It suffices to show that the missing component $f(R_1=0,X,Y)$ is identified.
We let $D(X)=A_2(X)-A_1(X)$ and note that
\begin{align*}
    \frac{f(R_1=1,X,Y)}{f(R_1=0,X,Y)}=\frac{f(R_1=1,X,Y)}{f(R_2=1,R_1=0,X,Y)}-\mathrm{exp}\{-D(X)\},
\end{align*}
where $f(R_1=1,X,Y)$, $f(R_1=1,X,Y)$ and $f(R_2=1,R_1=0,X,Y)$ are available from the observed data and the sampling weights, thus analogous to the proof of \citet[Theorem 1]{miao2025stableness}, the key is to identify $D(X)$.

We have
\begin{eqnarray}
   \nonumber &&\frac{f(R_2=0,R_1=0,X)}{f(R_2=1,R_1=0,X)} \\
	\nonumber& = & \int\left[\frac{\mathrm{exp}\{D(X)\}f(Y,R_1=1\mid X)}{f(Y,R_2=1,R_1=0\mid X)}-1\right]^{-1}f(Y\mid R_2=1,R_1=0,X)\odif Y\\
	& \equiv & L\{D(X)\}.\label{eq:dx}
\end{eqnarray}
This is an equation similar to Equation (7) in \citet{miao2025stableness}.
However, a key difference is that $X$ is also missing for nonrespondents here, and the quantity $f(R_2=0,R_1=0,X)$ in the left-hand side is not available only from the observed questionnaire data.
Nevertheless, by leveraging the covariates distribution $f(X)$,
we have $f(R_2=0,R_1=0,X)=f(X)-f(R_1=1,X)-f(R_2=1,R_1=0,X)$, which can be obtained from the observed data, the sampling weights and $f(X)$.
Thus, the left side of the above equation is available.
It is also easy to check that all quantities in the right side except for $D(X)$ are available.

Following the same argument as the proof of \citet[Theorem 1]{miao2025stableness},
for any fixed $x$ and any $D(x)$ such that
\begin{align*}
    \frac{f(Y,R_1=1\mid X=x)}{f(Y,R_2=1,R_1=0\mid X=x)}>\mathrm{exp}\{-D(X)\},
\end{align*}
$L\{D(x)\}$ is strictly decreasing in $D(x)$ because
\begin{eqnarray*}
\frac{\partial L\{D(x)\}}{\partial D(x)} = - \int \frac{\exp\{D(x)\}   \frac{f(Y, R_1=1\mid X=x)}{f(Y,R_2=1,R_1=0\mid X=x)}}{ \left[\exp\{D(x)\}  \frac{f(Y, R_1=1\mid X=x)}{f(Y,R_2=1,R_1=0\mid X=x)} - 1\right]^2} f(Y\mid R_2=1,R_1=0, X=x) dY <0.
\end{eqnarray*}
Thus,  for any fixed $x$ the solution to \eqref{eq:dx} is unique, and the identification of $D(X)$ is guaranteed by applying this argument to all $x$.

For the general case that $K\geq 3$, suppose that Assumption~\ref{assump1} holds for the first two calls.
By the above arguments, we can identify $f(R_1,R_2,X,Y)$ and then $f(X,Y)$.
Because $f(R_K=1,X,Y)$ is available from the observed-data distribution, identifying $f(X,Y)$ suffices to identify $f(R_K=0,X,Y)$, and thus the joint distribution $f(R_1,\ldots,R_K,X,Y)$ is identified.

Besides, identification can be achieved if Assumption~\ref{assump1} holds for any two given adjacent calls other than the first two.
To see this, suppose that the stableness of resistance holds for the $k$th and $k+1$th calls. 
By viewing the $k$th and $k+1$th calls as the first two calls in a subsurvey on nonrespondents from the   $k-1$th call (i.e., $R_{k-1}=0$) and by applying the proposed approach, 
we can identify $f(X,Y,R_k,R_{k+1}\mid R_{k-1}=0)$.
Noting that $f(X,Y,R_k,R_{k+1}\mid R_{k-1}=1)$ is available from the observed data, 
we can  identify $f(X,Y,R_{k-1},R_k,R_{k+1})$ and  thus $f(X,Y)$,
which suffices to  identify $f(X,Y,R_1,\ldots, R_K)$.

\subsection{Unbiasedness of the IPW and REG estimating equations}

\noindent \textit{Unbiasedness of the IPW estimating equations}:

We show that \eqref{eq:3}--\eqref{eq:6} are unbiased estimating equations for $(\alpha_1,\gamma,\alpha_2,\theta)$ if $\{A_1,\Gamma,A_2\}$ are correctly specified, and then the IPW estimator is consistent and asymptotically normal under standard regularity conditions for estimating equations \citep[see e.g.,][]{newey1994large,van2000asymptotic}.

Let $E(\cdot)$ denote the expectation with respect to the target population.
If $\{A_1,\Gamma,A_2\}$ are correctly specified, equations \eqref{eq:3}--\eqref{eq:5}
are unbiased estimating equations for $(\alpha_1,\alpha_2,\gamma)$ by noting
\begin{eqnarray}
   0 & = & E\left(\frac{R_1}{\pi_1}-1\mid X,Y\right),\label{eq:3e}\\
   0 & = & E\left\{\frac{R_2-R_1}{\pi_2}-(1-R_1)\mid X,Y\right\},\label{eq:4e}\\
   0 & = & E\left(\frac{R_2-R_1}{\pi_2}-\frac{1-\pi_1}{\pi_1}R_1\mid X,Y\right).\label{eq:5e}
\end{eqnarray}
Then $(\hat{\alpha}_{1,\ipw},\hat{\alpha}_{2,\ipw},\hat{\gamma}_{\ipw})$ converges to the truth $(\alpha_1,\alpha_2,\gamma)$ under standard regularity conditions for estimating equations \citep{newey1994large,van2000asymptotic}.

The probability limit of \eqref{eq:6} evaluated at $(\theta,\alpha_1,\alpha_2,\gamma)$ is
\begin{eqnarray*}
    E\left\{\frac{R_2m(X,Y;\theta)}{\pi_1+\pi_2(1-\pi_1)}\right\}=E\{m(X,Y;\theta)\}=0.
\end{eqnarray*}
Therefore, \eqref{eq:6} is an unbiased estimating equation for $\theta$ if $\{A_1,\Gamma,A_2\}$ are correctly specified.

\noindent \textit{Unbiasedness of the REG estimating equations}:

As before, $E(\cdot)$ denotes the expectation with respect to the target population.
If $\{A_1,\Gamma,f_2\}$ are correctly specified, equations \eqref{eq:7}--\eqref{eq:8}
are unbiased estimating equations for $(\alpha_1,\gamma,\beta)$ by noting
\begin{eqnarray*}
   &&E\left\{\left(\frac{R_1}{\pi_1}-R_2\right)U(X,Y)+R_2h_U(X;\beta,\gamma)\right\}-E\{h_U(X;\beta,\gamma)\}\\
   &=&E\left\{\left(\frac{R_1}{\pi_1}-1\right)U(X,Y)\right\} +E\left[(1-R_2)\left\{U(X,Y)-h_U(X;\beta,\gamma)\right\}\right]\\
   & = &0.
\end{eqnarray*}
Then $(\hat{\alpha}_{1,\reg},\hat{\gamma}_{\reg},\hat{\beta}_{\reg})$ converges to the truth $(\alpha_1,\gamma,\beta)$ under standard regularity conditions for estimating equations \citep{newey1994large,van2000asymptotic}.

The probability limit of \eqref{eq:9} evaluated at $(\theta,\alpha_1,\gamma,\beta)$ is
\begin{eqnarray*}
    &&E\left[R_2\left\{m(X,Y;\theta)-h_m(X;\theta,\beta,\gamma)\right\}\right]+E\{h_m(X;\theta,\beta,\gamma)\}\\
    &=&E\left[(R_2-1)\left\{m(X,Y;\theta)-h_m(X;\theta,\beta,\gamma)\right\}\right]+E\{m(X,Y;\theta)\}\\
    &=&0.
\end{eqnarray*}
Therefore, \eqref{eq:9} is an unbiased estimating equation for $\theta$ if $\{A_1,\Gamma,f_2\}$ are correctly specified.

\subsection{The double robustness of $\hat{\theta}_{\dr}$ in \eqref{eq:14}}\label{ssec:s3.3}

We show that \eqref{eq:10}--\eqref{eq:14} are unbiased estimating equations for $(\alpha_1,\gamma,\theta)$ if either $\{A_1,\Gamma,A_2\}$ or $\{A_1,\Gamma,f_2\}$ are correctly specified.
As before, $E(\cdot)$ denotes the expectation with respect to the target population.

\begin{itemize}
    
    \item If $\{A_1,\Gamma,A_2\}$ are correctly specified, then \eqref{eq:10}, \eqref{eq:11} and \eqref{eq:13} are unbiased estimating equations for $(\alpha_1,\alpha_2,\gamma)$ by noting \eqref{eq:3e}--\eqref{eq:5e}, and under standard regularity conditions for estimating equations \citep{newey1994large,van2000asymptotic}, $(\hat{\alpha}_{1,\dr},\hat{\gamma}_{\dr},\hat{\alpha}_{2,\dr})$ converges to the truth $(\alpha_1,\gamma,\alpha_2)$.
    Let $\beta^*$ denote the probability limit of $\hat{\beta}$, then the probability limit of \eqref{eq:14} evaluated at $(\theta,\gamma,\alpha_2,\beta^*)$ is
    \begin{eqnarray*}
        &&E\left[\left\{R_1+\frac{R_2-R_1}{\pi_2(X,Y;\alpha_2,\gamma)}\right\}\{m(X,Y;\theta)-h_m(X;\theta,\beta^*,\gamma)\}\right] +  E\{h_m(X;\theta,\beta^*,\gamma)\}\\
	& = & E\left\{\left(R_1+\frac{R_2-R_1}{\pi_2}\right)m(X,Y;\theta)\right\}-E\{h_m(X;\theta,\beta^*,\gamma)\}+E\{h_m(X;\theta,\beta^*,\gamma)\}\\
	& = & E\left\{\left(R_1+\frac{R_2-R_1}{\pi_2}\right)m(X,Y;\theta)\right\}\\
 & = & E\{m(X,Y;\theta)\}\\
 & = & 0.
    \end{eqnarray*}
    
Therefore, \eqref{eq:14} is an unbiased estimating equation for $\theta$ if $\{A_1,\Gamma,A_2\}$ are correctly specified.

    \item If $\{A_1,\Gamma,f_2\}$ are correctly specified, then \eqref{eq:12} is an unbiased estimating equation for $\beta$.
    Equations \eqref{eq:10} and \eqref{eq:13} are unbiased estimating equations for $(\alpha_1,\gamma)$ by noting \eqref{eq:3e} and
    \begin{eqnarray*}
    &&E\left[\left\{R_1-\frac{\pi_1(X,Y;\alpha_1,\gamma)}{1-\pi_1(X,Y;\alpha_1,\gamma)}\frac{R_2-R_1}{\pi_2(X,Y;\alpha_2^*,\gamma)}\right\}\{U(X,Y)-g_U(X;\beta)\}\right]\\
    &=&E\left[\left(R_1-\frac{\pi_1}{1-\pi_1}\frac{R_2-R_1}{\pi_2^*}\right)\{U(X,Y)-g_U(X;\beta)\}\right]\\
    &=&E\left[\left\{\pi_1-\frac{\pi_1}{1-\pi_1}\frac{(1-\pi_1)\pi_2}{\pi_2^*}\right\}\{U(X,Y)-g_U(X;\beta)\}\right]\\
    &=&E\left[\pi_1\pi_2\left(\frac{1}{\pi_2}-\frac{1}{\pi^*_2}\right)\{U(X,Y)-g_U(X;\beta)\}\right]\\
    && \text{from arguments in the proof of \citet[][Theorem 3]{miao2025stableness},}\\
    &=&0,
\end{eqnarray*}
where $\alpha_2^*$ is the probability limit of $\hat{\alpha}_{2,\dr}$.
    Then under standard regularity conditions for estimating equations, $(\hat{\alpha}_{1,\dr},\hat{\gamma}_{\dr},\hat{\beta}_{\dr})$ are consistent and asymptotically normal.

    The probability limit of \eqref{eq:14} evaluated at $(\theta,\gamma,\alpha_2^*,\beta)$ is
    \begin{eqnarray*}
        &&E\left[\left\{R_1+\frac{R_2-R_1}{\pi_2(X,Y;\alpha_2^*,\gamma)}\right\}\{m(X,Y;\theta)-h_m(X;\theta,\beta,\gamma)\}\right] +  E\{h_m(X;\theta,\beta,\gamma)\}\\
	& = & E\left[\left(R_1+\frac{R_2-R_1}{\pi^*_2}-1\right)\{m(X,Y;\theta)-h_m(X;\theta,\beta,\gamma)\}\right]+E\{m(X,Y;\theta)\}\\
 & = &E\left[\left\{\pi_1+\frac{(1-\pi_1)\pi_2}{\pi^*_2}-1\right\}\{m(X,Y;\theta)-h_m(X;\theta,\beta,\gamma)\}\right]+E\{m(X,Y;\theta)\}\\
 & = &E\left[\left\{(1-\pi_1)\pi_2\left(\frac{1}{\pi^*_2}-\frac{1}{\pi_2}\right)\right\}\{m(X,Y;\theta)-h_m(X;\theta,\beta,\gamma)\}\right]+E\{m(X,Y;\theta)\}\\
 && \text{from arguments in the proof of \citet[][Theorem 3]{miao2025stableness},}\\
	& = & E\{m(X,Y;\theta)\}\\
 & = & 0.
    \end{eqnarray*}
\end{itemize}
Therefore, \eqref{eq:14} is an unbiased estimating equation for $\theta$ if $\{A_1,\Gamma,f_2\}$ are correctly specified.

In summary, \eqref{eq:10}--\eqref{eq:14} are unbiased estimating equations for $(\alpha_1,\gamma,\theta)$ if either $\{A_1,\Gamma,A_2\}$ or $\{A_1,\Gamma,f_2\}$ are correctly specified, and under standard regularity conditions for estimating equations, $(\hat{\alpha}_{1,\dr},\hat{\gamma}_{\dr},\hat{\theta}_{\dr})$ are consistent and asymptotically normal if either $\{A_1,\Gamma,A_2\}$ or $\{A_1,\Gamma,f_2\}$ are correctly specified.

\subsection{Proof of Theorem~\ref{theorems.1}}\label{ssec:s3.4}

As shown in Section~\ref{ssec:s3.3}, $(\hat\alpha_{1,\dr},\hat\gamma_\dr)$ are  consistent and asymptotically normal if 
$A_1(\alpha_1),\Gamma(\gamma)$ and at least one of   $A_2(\alpha_2), f_2(\beta)$ are correct.
We   show that \eqref{eq:drs.1} is unbiased for estimating $\gamma_K$ if further $\Gamma_K(\gamma_K)$ and at least one of $A_K(\alpha_K),f_K(\beta_K)$ is correct.

If $\Gamma_K(\gamma_K)$ and $A_K(\alpha_K)$ are correct, then \eqref{eq:drs.1} is unbiased for estimating $(\alpha_K,\gamma_K)$
by noting that 
\begin{equation}
E \left\{\frac{R_K - R_2}{ \pi_K(\alpha_K,\gamma_K)} + R_2 -  \frac{R_1}{\pi_1} \mid X,Y \right\} =0.
\end{equation}
If $\Gamma_K(\gamma_K)$ and $f_K(\beta_K)$ are correct, then \eqref{eq:regs.1}  is unbiased for estimating $\beta_K$.
Letting $\alpha_K^*$ be the probability limit of $\hat\alpha_{K,\dr}$ and $\pi_K^*=\pi_K(\alpha_K^*,\gamma_K)$,
then the probability limit of  \eqref{eq:drs.1} evaluated at $(\alpha_K^*, \gamma_K, \alpha_1, \gamma, \beta_K)$ is 
\begin{eqnarray*}
&&E\left[ \left\{\frac{R_K - R_2}{ \pi_K^*} + R_2 -  \frac{R_1}{\pi_1} \right\} \cdot
\left\{U(X,Y) -  E(U(X,Y)\mid X, R_K=0;\beta_K,\gamma_K) \right\}\right]\\
&=&  E\left[\left\{\frac{(1-p_2)\pi_K}{\pi_K^*}  + p_2 -1 \right\}
\{U(X,Y) -  E(U(X,Y)\mid X, R_K=0)\} \right],\\
&=& E\left[\left\{ (1-p_2)\pi_K \left( \frac{1}{\pi_K^*}-\frac{1}{\pi_K}\right) \right\}
\{U(X,Y) -  E(U(X,Y)\mid X, R_K=0)\} \right],\\
&&\text{letting   $g=(1/\pi_K^* - 1/\pi_K)/e^{-\Gamma_K}$,}\\
&=& E\left[  (R_K-R_2)e^{-\Gamma_K}   g  \{U(X,Y) -  E(U(X,Y)\mid X, R_K=0)\} \right],\\
&&\text{noting that  $g$ is a function   only of $X$,}\\
&=& E\left[  E \left\{(R_K-R_2) e^{-\Gamma_K} gU(X,Y)  \mid X\right\} -   E(gU(X,Y)\mid X, R_K=0)\cdot E \left\{(R_K - R_2) e^{-\Gamma_K} \mid X\right\} \right]\\
&=& 0.
\end{eqnarray*}
Therefore,   \eqref{eq:drs.1} is unbiased for estimating $\gamma_K$ if $A_1(\alpha_1),\Gamma(\gamma), \Gamma_K(\gamma_K)$, at least one of   $A_2(\alpha_2), f_2(\beta)$,
and at least one of $A_K(\alpha_K),f_K(\beta_K)$ are correctly specified.
An analogous argument shows that \eqref{eq:drs.2} is unbiased for estimating $\theta$ if $A_1(\alpha_1),\Gamma(\gamma), \Gamma_K(\gamma_K)$, at least one of   $A_2(\alpha_2), f_2(\beta)$,
and at least one of $A_K(\alpha_K),f_K(\beta_K)$ are correctly specified.

\section{Additional simulation results}\label{sec:s4}

\subsection{Simulations for sensitivity analysis against the violation of the stableness of resistance assumption}\label{ssec:simu-sens}

We assess robustness   of the proposed estimators  in situations where the stableness of resistance assumption does not hold.  
The difference $\Delta$ between the log odds ratios in two calls is used as the sensitivity parameter capturing the degree of violation of the assumption.
For each value of $\Delta$ among $\pm(0,0.1,0.2)$, we generate and analyze data according to the following 
models and replicate 1000 simulations at sample size 5000.
\begin{table}[!ht]
\centering
\caption{Data generating and estimation models for sensitivity analysis}
\begin{tabular}{c}
\toprule
Data generating model\\
$\pi_1=\expit(\alpha_1^\T X + \gamma_1 Y),\pi_2=\expit\{\alpha_2^\T  X + (\gamma_1+\Delta) Y\}, f_2(Y=1\mid X)=\expit(\beta^\T X)$\\
$\alpha_1^\T =(-1, 0.5, 0.2) $, $\alpha_2^\T =(-0.5, 0.5, 0.2) $, $\beta^\T =(-0.5, 0.5, 0.5) $\\
$\gamma_1=0.5$\\ 
Working model for estimation\\
$\pi_1=\expit(\alpha_1^\T X + \gamma Y),\pi_2=\expit(\alpha_2^\T X + \gamma Y), f_2(Y=1\mid X)=\expit(\beta^\T X)$\\
\bottomrule
\end{tabular}
\end{table}

\begin{figure}[!ht]
\centering
\graphicspath{{figures/}}
\includegraphics[scale=0.65]{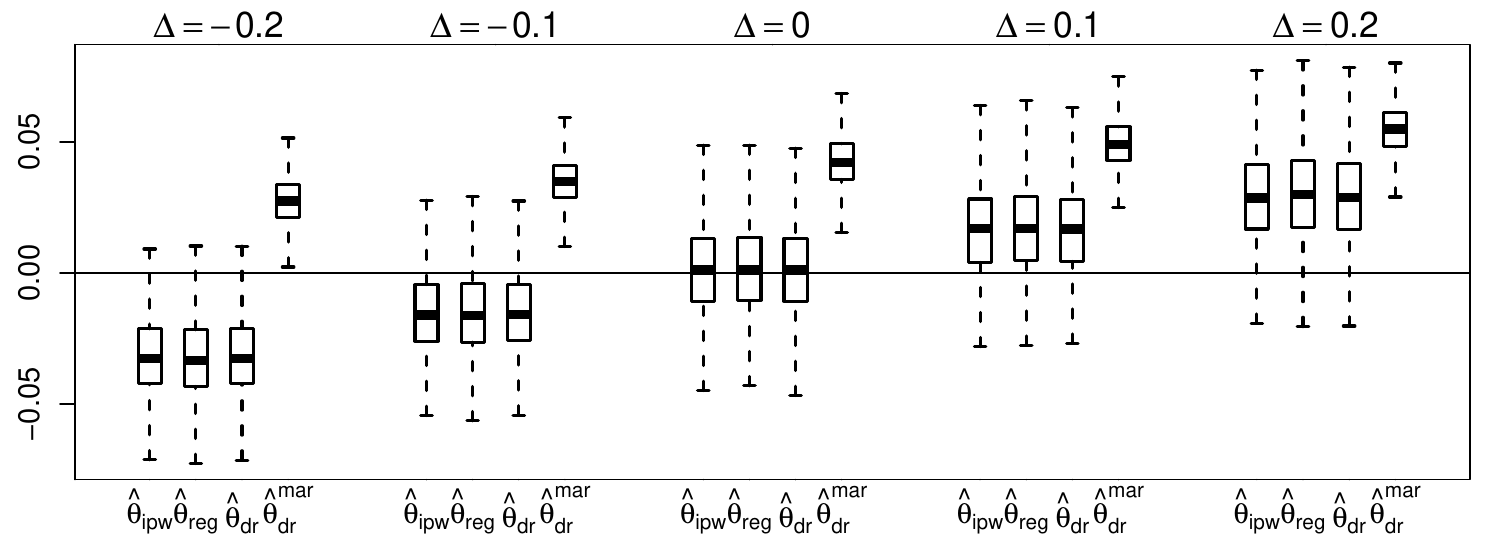}
\caption{Bias   of   estimators of $\theta$ at different  levels of $\Delta$ in the simulation. The estimator $\hat \theta_\dr^{\mathrm{mar}}$ is the AIPW estimator assuming MAR.}\label{fig:sensitivity}     
\end{figure}

 Figure~\ref{fig:sensitivity} shows the bias of estimators.
The proposed estimators   generally overestimate the outcome mean  when $\Delta>0$ and underestimate when $\Delta<0$.
The bias  of the three proposed estimators assuming stableness of resistance  is small when the sensitivity parameter varies within a moderate range, 
but the bias increases with the magnitude of $\Delta$ and could become large if the stableness of resistance assumption is severely violated.

\subsection{Simulations for the continuous outcome setting}\label{ssec:simu-con}

We evaluate the performance of the proposed estimators in the case of a continuous outcome via simulation studies.
Let $X = (1,X_a,X_b)^\T$ and $\widetilde{X} = (1,X_a^2,X_b^2)^\T$
with $X_a,X_b$ independent following a uniform distribution $\operatorname{Unif}(-1,1)$.
Table~\ref{tbl:s.2} presents settings   for data generation and estimation in the simulation.
The samples of    $(X,Y)$  are dropped for $R_2=0$ and the marginal  distribution of $X$ is given for estimation.

\begin{table}[!ht]
\caption{Models for data generation and estimation in the simulation}
\label{tbl:s.2}
    \centering
    \begin{tabular}{cccccccc}
    \toprule
  &  &\multicolumn{4}{c}{Data generating model for continuous outcome}\\
  \cmidrule{3-6}
  &  &\multicolumn{4}{c}{$\pi_1=\expit(\alpha_1^\T X+\gamma Y), \pi_2=\expit(\alpha_2^\T W_1+\gamma Y),f_2(Y\mid X)\sim N(\beta^\T W_2, \sigma^2)$}\\
  \cmidrule{3-6}
  &  &\multicolumn{4}{c} {Four scenarios with different choices of $(W_1, W_2)$}\\
  \midrule
    &     Settings     &TT &FT &TF &FF   \\ 
    \midrule
   &$(W_1,W_2)$ &$(X,X)$ &$(\widetilde{X},X)$ &$(X,\widetilde{X})$ &$(\widetilde{X},\widetilde{X})$\\
&$\alpha_1^\T$ & $(0, 0.6, 0.5)$ & $(-0.35, -0.5, 0.7)$ & $(-1, 1, -0.1)$ & $(-0.3, -0.5, 1)$ \\
&$\alpha_2^\T$ & $(1.4, -0.5, 0.2)$ & $(-0.5, 1.8, 1)$ & $(0.5, 1, -0.1)$ & $(-0.4, 0.8, 0)$ \\
&$\beta^\T$ & $(0.6, 1.0, 0.3)$ & $(-0.8, 5, 3.5)$ & $(-0.5, 5, -1)$ & $(-1.5, 4, 3)$ \\
&$\gamma$ & $0.13$ & $0.12$ & $0.5$ & $0.25$ \\
&$\sigma^2$ & $3$ & $2$ & $0.4$ & $0.25$ \\
&  &\multicolumn{4}{c}{Working model for estimation}\\
&  &\multicolumn{4}{c}{$\pi_1=\expit(\alpha_1^\T X+\gamma Y), \pi_2=\expit(\alpha_2^\T X+\gamma Y),f_2(Y\mid X)\sim N(\beta^\T X, \sigma^2),$}\\
\bottomrule
    \end{tabular}
\end{table}

All working models are correctly specified in Scenario (TT). However, in Scenarios (TF) and (FF), the working model for the second-call outcome model is misspecified, and in Scenarios (FT) and (FF), the working model for the second-call baseline propensity score model is misspecified.
We implement the proposed IPW, REG, and  DR  methods to estimate the outcome mean, i.e.
the solution to  $E\{m(X,Y;\theta)\}=E(Y-\theta)=0$.

We simulate 1,000 replicates for each scenario with  sample size of 5,000 in each replicate. 
We set the functions $V_1(x)=V_2(x)=x$, and $U(x,y)=y$ for the IPW estimator; $U(x,y)=(x^\T,y)^\T$ for the REG estimator;  $V_1(x)=V_2(x)=x$, and $U(x,y)=y$ for the DR estimator.
For comparison, we also implement standard  estimators $(\hat \theta_{\ipw}^{\mathrm{mar}},\hat \theta_{\reg}^{\mathrm{mar}},\hat \theta_{\dr}^{\mathrm{mar}})$ that are IPW, REG, DR analogs based on MAR, with  the number of callbacks included as an additional covariate, and all covariates are fully available.
The simulation results are summarized with  boxplots  of estimation bias in Figure~\ref{fig:s.4} for the outcome mean 
 $\theta$ and Figure~\ref{fig:s.5} for the odds ratio
parameter $\gamma$. 

As expected, the  three proposed estimators have little bias in Scenario
(TT) where  all working models are correctly specified.
However, the IPW and REG estimators have substantial bias when the second-call baseline propensity score model $A_2(x;\alpha_2)$ and the second-call outcome model $f_2(y\mid x;\beta)$ are misspecified, respectively. In contrast, the DR estimator continues to exhibit little bias in these scenarios.
These results demonstrate the double robustness of $(\hat \theta_{\dr}, \hat \gamma_{\dr})$ against misspecification of either $A_2(x;\alpha_2)$ or $f_2(y\mid x;\beta)$. 
In Scenario (FF), where both $A_2(x;\alpha_2)$ and $f_2(y\mid x;\beta)$ are misspecified, all three proposed estimators lead to biased estimates.
Besides, the three standard MAR estimators have large bias in all four scenarios, even if the number of callbacks is included as a covariate and all covariates are fully available.

We compute the variance of the proposed estimators and construct 95\% confidence intervals based on the normal approximation of their distributions. 
We then assess the coverage rate of the confidence intervals and summarize the results in Table~\ref{tbl:s.3}. 
In Scenarios (TT), (FT) and (TF), the 95\% confidence interval  based on the DR estimator  has a coverage rate that is very close to the nominal  level of 0.95. 
However, in the case of the corresponding working model being misspecified,  the 95\% confidence intervals based on the IPW and REG estimators  have undersized coverage rates.

In summary, we recommend the proposed methods for nonresponse adjustment with callback data and the marginal distribution of missing covariates, and we suggest using several different working models to improve robustness.

\begin{figure}[!ht]
\graphicspath{{figures/Figure_S4/}}
\centering
\includegraphics[scale=0.45]{TT.pdf}
\includegraphics[scale=0.45]{FT.pdf}
\includegraphics[scale=0.45]{TF.pdf}
\includegraphics[scale=0.45]{FF.pdf}
\caption{Bias of estimators of $\theta$ in the continuous outcome simulation.} \label{fig:s.4}
\end{figure}
\begin{figure}[!ht]
\graphicspath{{figures/Figure_S5/}}
\centering
\includegraphics[scale=0.45]{TT.pdf}
\includegraphics[scale=0.45]{FT.pdf}
\includegraphics[scale=0.45]{TF.pdf}
\includegraphics[scale=0.45]{FF.pdf}
\caption{Bias of estimators of $\gamma$ in the continuous outcome simulation.}
\label{fig:s.5}
\end{figure}

 \begin{table}[!ht] 
\caption{Coverage rate of $95\%$ confidence interval in the continuous outcome simulation} \label{tbl:s.3} 
\centering
\begin{tabular}{ccccccccccc}
\toprule  &\multicolumn{6}{c}{$\theta$} && \multicolumn{3}{c}{$\gamma$}\\ 
\cmidrule{2-7}\cmidrule{9-11}
Scenarios& IPW & REG  & DR & IPW$_\mathrm{mar}$ & REG$_\mathrm{mar}$  & DR$_\mathrm{mar}$ & &IPW  & REG  & DR   \\
\midrule
TT &{0.958}  & {0.957} & {0.956} &{0.109} &{0.362} &{0.363} & &{0.953} & {0.952} & {0.958} \\ 
FT & {0.446} & {0.959} & {0.954} &{0.000}&{0.442}&{0.445} & &{0.334} &  {0.959}  & {0.949} \\
TF & {0.957}  & {0.131} & {0.957} &{0.000}&{0.000}&{0.000} & &{0.946} & {0.495} & {0.951}    \\
FF & {0.286} & {0.464} & {0.367} &{0.000}&{0.000}&{0.000}  & &{0.474} & {0.623}  & {0.681}  \\
\bottomrule
\end{tabular} 
\end{table}

\section{Additional data analysis results}\label{sec:s5}
\subsection{Demographic covariates distribution obtained from the census data}\label{ssec:demo-dist}
Tables~\ref{tbl:demo-dist-i} and~\ref{tbl:demo-dist-ii}  show the demographic covariates distribution obtained from the census data\footnote{Source:  U.S. Census Bureau, Current Population Survey, 2020 Annual Social and Economic Supplement.
See \href{https://www.census.gov/data/tables/2020/demo/educational-attainment/cps-detailed-tables.html}{\texttt{https://www.census.gov/data/tables/2020/demo/educational-attainment/cps-detailed-tables.html}} for more details.}, with the numbers of people in each subgroup defined by demographic covariates in the NRFU observed data.
The numbers of column ``Census Data'' are numbers in thousands. 
The column ``NRFU'' corresponds to the numbers of people in each subgroup defined by demographic covariates in the NRFU observed data.

\begin{table}[!ht]
\caption{Demographic covariates frequencies I}
\label{tbl:demo-dist-i}
\centering
\small
\renewcommand{\arraystretch}{0.8}
\begin{tabular}{lllllcc}
\toprule
\textbf{Race} & \textbf{Ethnicity} & \textbf{Gender} & \textbf{Age} & \textbf{Educational Attainment} & \textbf{Census Data} & \textbf{NRFU}\\
\midrule
White & Hispanic & Male & 18--29 & High school or less & 2744 & 1\\
White & Hispanic & Male & 18--29 & Some college & 1862 & 1\\
White & Hispanic & Male & 18--29 & Bachelor's degree or above & 653 & 1\\
White & Hispanic & Male & 30--59 & High school or less & 6568 & 5\\
White & Hispanic & Male & 30--59 & Some college & 2105 & 5\\
White & Hispanic & Male & 30--59 & Bachelor's degree or above & 1868 & 12\\
White & Hispanic & Male & 60+ & High school or less & 1784 & 11\\
White & Hispanic & Male & 60+ & Some college & 535 & 3\\
White & Hispanic & Male & 60+ & Bachelor's degree or above & 545 & 4\\
White & Hispanic & Female & 18--29 & High school or less & 2325 & 1\\
White & Hispanic & Female & 18--29 & Some college & 1904 & 4\\
White & Hispanic & Female & 18--29 & Bachelor's degree or above & 825 & 3\\
White & Hispanic & Female & 30--59 & High school or less & 5509 & 6\\
White & Hispanic & Female & 30--59 & Some college & 2368 & 23\\
White & Hispanic & Female & 30--59 & Bachelor's degree or above & 2331 & 18\\
White & Hispanic & Female & 60+ & High school or less & 2245 & 4\\
White & Hispanic & Female & 60+ & Some college & 592 & 4\\
White & Hispanic & Female & 60+ & Bachelor's degree or above & 580 & 1\\
White & Non-Hispanic & Male & 18--29 & High school or less & 5793 & 15\\
White & Non-Hispanic & Male & 18--29 & Some college & 4755 & 17\\
White & Non-Hispanic & Male & 18--29 & Bachelor's degree or above & 3590 & 12\\
White & Non-Hispanic & Male & 30--59 & High school or less & 11412 & 59\\
White & Non-Hispanic & Male & 30--59 & Some college & 9569 & 77\\
White & Non-Hispanic & Male & 30--59 & Bachelor's degree or above & 15634 & 143\\
White & Non-Hispanic & Male & 60+ & High school or less & 9430 & 84\\
White & Non-Hispanic & Male & 60+ & Some college & 6773 & 76\\
White & Non-Hispanic & Male & 60+ & Bachelor's degree or above & 10347 & 141\\
White & Non-Hispanic & Female & 18--29 & High school or less & 4066 & 17\\
White & Non-Hispanic & Female & 18--29 & Some college & 5030 & 16\\
White & Non-Hispanic & Female & 18--29 & Bachelor's degree or above & 4482 & 28\\
White & Non-Hispanic & Female & 30--59 & High school or less & 9087 & 50\\
White & Non-Hispanic & Female & 30--59 & Some college & 10503 & 105\\
White & Non-Hispanic & Female & 30--59 & Bachelor's degree or above & 17969 & 167\\
White & Non-Hispanic & Female & 60+ & High school or less & 11732 & 111\\
White & Non-Hispanic & Female & 60+ & Some college & 8491 & 121\\
White & Non-Hispanic & Female & 60+ & Bachelor's degree or above & 9590 & 145\\
\midrule
Total &&&&&195596 & 1491\\
\bottomrule
\end{tabular}
\end{table}

\begin{table}[!ht]
\caption{Demographic covariates frequencies II}
\label{tbl:demo-dist-ii}
\centering
\small
\renewcommand{\arraystretch}{0.8}
\begin{tabular}{lllllcc}
\toprule
\textbf{Race} & \textbf{Ethnicity} & \textbf{Gender} & \textbf{Age} & \textbf{Educational Attainment} & \textbf{Census Data} & \textbf{NRFU}\\
\midrule
Nonwhite & Hispanic & Male & 18--29 & High school or less & 379 & 0\\
Nonwhite & Hispanic & Male & 18--29 & Some college & 262 & 0\\
Nonwhite & Hispanic & Male & 18--29 & Bachelor's degree or above & 117 & 1\\
Nonwhite & Hispanic & Male & 30--59 & High school or less & 597 & 2\\
Nonwhite & Hispanic & Male & 30--59 & Some college & 230 & 0\\
Nonwhite & Hispanic & Male & 30--59 & Bachelor's degree or above & 300 & 2\\
Nonwhite & Hispanic & Male & 60+ & High school or less & 221 & 3\\
Nonwhite & Hispanic & Male & 60+ & Some college & 46 & 1\\
Nonwhite & Hispanic & Male & 60+ & Bachelor's degree or above & 74 & 0\\
Nonwhite & Hispanic & Female & 18--29 & High school or less & 334 & 3\\
Nonwhite & Hispanic & Female & 18--29 & Some college & 241 & 5\\
Nonwhite & Hispanic & Female & 18--29 & Bachelor's degree or above & 135 & 0\\
Nonwhite & Hispanic & Female & 30--59 & High school or less & 642 & 8\\
Nonwhite & Hispanic & Female & 30--59 & Some college & 371 & 4\\
Nonwhite & Hispanic & Female & 30--59 & Bachelor's degree or above & 327 & 5\\
Nonwhite & Hispanic & Female & 60+ & High school or less & 224 & 1\\
Nonwhite & Hispanic & Female & 60+ & Some college & 89 & 5\\
Nonwhite & Hispanic & Female & 60+ & Bachelor's degree or above & 66 & 3\\
Nonwhite & Non-Hispanic & Male & 18--29 & High school or less & 2653 & 6\\
Nonwhite & Non-Hispanic & Male & 18--29 & Some college & 2114 & 6\\
Nonwhite & Non-Hispanic & Male & 18--29 & Bachelor's degree or above & 1394 & 1\\
Nonwhite & Non-Hispanic & Male & 30--59 & High school or less & 4656 & 12\\
Nonwhite & Non-Hispanic & Male & 30--59 & Some college & 2864 & 11\\
Nonwhite & Non-Hispanic & Male & 30--59 & Bachelor's degree or above & 5049 & 35\\
Nonwhite & Non-Hispanic & Male & 60+ & High school or less & 2451 & 18\\
Nonwhite & Non-Hispanic & Male & 60+ & Some college & 1223 & 10\\
Nonwhite & Non-Hispanic & Male & 60+ & Bachelor's degree or above & 1548 & 20\\
Nonwhite & Non-Hispanic & Female & 18--29 & High school or less & 2322 & 5\\
Nonwhite & Non-Hispanic & Female & 18--29 & Some college & 2260 & 2\\
Nonwhite & Non-Hispanic & Female & 18--29 & Bachelor's degree or above & 1846 & 10\\
Nonwhite & Non-Hispanic & Female & 30--59 & High school or less & 4329 & 17\\
Nonwhite & Non-Hispanic & Female & 30--59 & Some college & 3642 & 28\\
Nonwhite & Non-Hispanic & Female & 30--59 & Bachelor's degree or above & 6446 & 35\\
Nonwhite & Non-Hispanic & Female & 60+ & High school or less & 3323 & 18\\
Nonwhite & Non-Hispanic & Female & 60+ & Some college & 1742 & 20\\
Nonwhite & Non-Hispanic & Female & 60+ & Bachelor's degree or above & 2003 & 19\\
\midrule
Total &&&&&56520 & 316\\
\bottomrule
\end{tabular}
\end{table}

\subsection{Heterogeneity of response}\label{ssec:inta}
In addition, there might exist heterogeneous effect of the design covariates  on the response and heterogeneous nonignorability across different demographic groups.
We conduct additional analysis to investigate such heterogeneity by including interaction terms between design covariates and demographic covariates, and interaction terms   between the outcome and demographic covariates  in the propensity score models. 
In particular, we examine the interaction of the visible cash incentive with gender and age, and the interaction  between the outcome and gender.
The extended covariates $X=(1, \textit{race}, \textit{ethnicity}, \textit{gender},  \textit{age2}, \textit{age3}, \textit{edu2}, \textit{edu3}, \textit{m1sent},\\\textit{version}, \textit{title}, \textit{incvis}, \textit{incvis:gender}, \textit{incvis:age2}, \textit{incvis:age3})^\T$.
We implement the proposed SOR-based doubly robust estimator, with parametric  working models $\pi_1(x,y;\alpha_1,\gamma)=\operatorname{expit}(\alpha_1^\T x+\gamma^\T x_{\rm gender}y)$,  $\pi_2(x,y;\alpha_2,\gamma)=\operatorname{expit}(\alpha_2^\T x+\gamma^\T x_{\rm gender}y)$ and $f_2(y=1\mid x;\beta)=\operatorname{expit}(\beta^\T x)$, where $x_{\rm gender}=(1,\textit{gender})^\T$.

Table~\ref{tbl:s.ita} reports the point estimates and 95\%
 confidence intervals for covariates coefficients and the odds ratio parameter $\gamma$ in the propensity score models.
  
 \begin{table}[!ht] 
\caption{Point estimates and  95\% confidence intervals (C.I.) for  covariates coefficients and the odds ratio parameter in propensity score models}
\label{tbl:s.ita}
\centering
\begin{tabular}{l l l l l  }
\toprule  & \multicolumn{2}{c}{\centering \textbf{First contact}} & \multicolumn{2}{c}{\centering \textbf{Second contact}}  \\
\cmidrule{2-5}
{Covariate} & {Estimate} & {95\% C.I.} & {Estimate} &  95\%  C.I.\\
\midrule
\textit{intercept}   &  -2.347 &          (-3.117,     -1.577)  &  \hspace{0.1em}-2.714 & (-3.515,     -1.913) \\
\textit{m1sent}:  advance postcard   &  -0.230 &          (-0.578,      0.118)   &  \hspace{0.1em}-0.251 & (-0.780,      0.278) \\
\textit{version}:  on page 1   &  \hspace{0.1em} 0.230 &          (-0.069,      0.528)   &  \hspace{0.3em}0.007 & (-0.538,      0.552) \\
\textit{title}: long     &     -0.049   &  (-0.345,      0.247)        &   -0.097& (-0.654,      0.460)\\
\textit{incvis}: visible     &  \hspace{0.1em} 0.062   &        (-0.702,      0.826)
     & \hspace{0.3em}-0.046 & (-1.153,      1.061)\\
\textit{race}: nonwhite     &   -0.163  & (-0.564,      0.239)        &  \hspace{0.1em}-0.107 & (-0.647,      0.433)\\
\textit{ethnicity}: Hispanic & -0.125 & (-2.213, 1.964)
& -0.790 & (-2.994, 1.414)\\
\textit{gender}: male     &  -0.790     &  (-1.224,     -0.357)         &  -0.893 & (-1.513,     -0.273) \\
\textit{age2}: 30--59     & \hspace{0.1em} 0.825    &      (0.212,      1.439)      & \hspace{0.3em}1.016 & (0.277,      1.756)\\
\textit{age3}: 60+     &  \hspace{0.3em}1.729    &         (1.010,      2.448)     & \hspace{0.3em}2.085 & (1.125,      3.044)  \\
\textit{edu2}: some college     &    \hspace{0.1em} 0.296    &   (-0.302,      0.894)       & \hspace{0.3em}0.706 & (0.104,      1.308)\\
\textit{edu3}: Bachelor's degree or above     &    \hspace{0.1em} 0.397    &   (-0.362,      1.157)       & \hspace{0.3em}0.635 & (-0.074,      1.344)\\
\textit{incvis:gender} &\hspace{0.1em} 0.676 & (0.108,      1.244)& \hspace{0.2em} 0.810 & (-0.021,      1.642)\\
\textit{incvis:age2} &-0.288 & (-1.119, 0.543) & -0.389 &(-1.516, 0.737)\\
\textit{incvis:age3} &-0.538 & (-1.408, 0.333) & -0.532 &(-1.975, 0.910)\\
\midrule
\textit{gamma}: intercept &\hspace{0.1em} 1.632 &(0.526,      2.738) && \\
\textit{gamma}: gender &-0.717 &(-4.350,      2.916) && \\
\bottomrule
\end{tabular}
\end{table}

The analysis results in Table~\ref{tbl:s.ita} show a  statistically significant interaction effect of  the  visible cash incentive and gender on the response to the first call (point estimate 0.676, 95\% C.I. (0.108, 1.244)), 
which suggests that the visible cash incentive could increase the  response rate more effectively among men than among women.
This finding aligns with evidence that monetary rewards  tend to motivate men more  than women in tasks \citep[e.g.,][]{spreckelmeyer2009anticipation,czap2018comparing}, 
suggesting proper cash incentives could partially compensate for overrepresentation in gender-imbalanced samples.
Our analysis does not show strong evidence for  the interaction between the the  visible cash incentive and age, or the interaction between gender and the outcome.

In principle, it is possible to  further extend our  model to include additional interactions terms to study the potential heterogeneous effect. 
However, this would dramatically  increase the  model complexity,  lead to unstable estimates, and require a large sample size.
While acknowledging  that the investigation of additional heterogeneity is of important interest and could be addressed properly, we leave this to future work when this is supported with sufficient samples.

\subsection{Discussion of the applicability of the RRI design in the NRFU data}\label{ssec:rri}

While our callback-based SOR approach successfully corrects the nonresponse bias in the NRFU survey, alternative frameworks such as randomized response instruments \citep[RRI,][]{bailey2024polling,bailey2025countering} can offer other effective means when callbacks are not available. 
RRI is a useful tool for mitigating nonignorable nonresponse---if the RRIs are strongly associated with the response, 
the RRI model would lead to a fairly dramatic lowering of the nonresponse bias.
We note that   the NRFU  study indeed encompasses a randomized response instrument design with several randomized design features, including the advance postcard (\textit{m1sent}, sent or not), the questionnaire version (\textit{version}, political content on page 2 or on page 1), study title (\textit{title}, short or long), and prepaid incentive presentation (\textit{invis}, visible or not). 
In principle the RRI approach could also be applied;
however,  as shown in our analysis below, in the NRFU study the association between  the randomized design features and the response  is weak, and thus, we do not implement the RRI method. 
To show this, we fit the simple logistic regression model of $R_2$ on $X_2$:
\[f(R_2=1\mid X_2)=\expit(\beta_0+\beta_1\textit{m1sent}+\beta_2\textit{version}+\beta_3\textit{title}+\beta_4\textit{incvis}).\]
The $p$-value of the likelihood ratio test for the null hypothesis: $\beta_1=\beta_2=\beta_3=\beta_4=0$ is 0.418, which suggests that the association between  the randomized design features and the response  is weak.

For future studies, both RRI and SOR frameworks should be considered, as long as they are suited to the study, to strengthen the data analysis with nonignorable nonresponse.
\putbib
\end{bibunit}

\end{document}